\documentclass{article}

\usepackage{arxiv}

\usepackage[utf8]{inputenc} 
\usepackage[T1]{fontenc}    
\usepackage{hyperref}       
\usepackage{url}            
\usepackage{booktabs}       
\usepackage{nicefrac}       
\usepackage{microtype}      
\usepackage{lipsum}		

\usepackage{dblfloatfix}
\usepackage{cite}
\usepackage{amsmath,amssymb,amsfonts}
\usepackage{algorithmic}
\usepackage{graphicx}
\usepackage{textcomp}
\usepackage{caption,subcaption}
\usepackage{array}
\usepackage{epsfig}
\usepackage{epstopdf}
\usepackage{multicol}
\usepackage[ruled,vlined,linesnumbered,noresetcount]{algorithm2e}
\usepackage{mathtools}
\usepackage{cuted}
\usepackage{blindtext}
\usepackage{url}
\usepackage{soul}
\usepackage{color}
\usepackage{longtable}
\usepackage{enumitem}
\usepackage{multirow}

\title{Employing Game Theory and TDMA Protocol to Enhance Security and Manage Power Consumption in WSNs-based Cognitive Radio}

\author{
  Mohamed S. Abdalzaher$^{1,2}$\thanks{$^2$National Research Institute of Astronomy and Geophysics, Seismology Department, 11421, Egypt} $\;$ and Osamu Muta$^1$ \\
  $^1$Center for Japan-Egypt Cooperation\\ in Science and Technology, \\Kyushu University, \\744 Motooka, Nishi-ku, Fukuoka-shi 819-0395, Japan
  \texttt{mohamed.abdelzaher@ejust.edu.eg} \\ }
  
\begin{document}
\maketitle

\begin{abstract}
The rapid development of wireless sensor networks (WSNs) is the significant incentive to contribute in the vulnerable applications such as cognitive radio (CR). This paper proposes a Stackelberg game approach to enhance the WSN-based CR security against the spectrum sensing data falsification (SSDF) attack and conserve the consequent lost power consumption. The attack aims to corrupt the spectrum decision by imposing interference power to the delivered reports from the sensor nodes (SNs) to the fusion center (FC) to make a protection level below a specific threshold. The proposed model utilizes the intelligent Stackelberg game features along with the matched filter (MF) to maximize the number of protected reports sent by the SNs to the FC leading to accurate decision of the spectrum status. Furthermore, the TDMA protocol is utilized to resolve the complexity of employing MF for the spectrum detection to avoid the collision between the delivered reports. The proposed model aims to enhance the number of correctly received reports at the FC, and hence manage the lost energy of reports retransmission due to the malicious attack effect. Moreover, the model can conserve the lost power of the failure communication attempts due to the SSDF attack impact. Simulation results indicate the improved performance of the proposed protection model along with the MF over the six different environments against the SSDF attack  as compared to two defense schemes, namely, random and equal weight defense strategies. 
\end{abstract}



\textbf{\textit{Key}words} Wireless Sensor Networks, Cognitive Radio, Game Theory, Threats Mitigation, Power Conservation.



\section{Introduction}
\label{sec:introduction}

With the dramatic lack of spectrum resources, the intelligent communication methodologies have flourished such as cognitive radio (CR) to resolve this issue. In the literature context, the spectrum quality-of-service (QoS) management, energy conservation for which the packet size optimization was one of the key roles, and security mitigation rise among most of the exerted efforts. Various CR paradigms were studied focusing on the spectrum utilization depending on the statistical QoS to support the real-time applications at the secondary users (SUs) \cite{wang2019unified, anwar2016effective,shah2013cross,alshamrani2011qos,   arafa2013feedback, guirguis2015primary}. Moreover, the SUs are used to support the primary feedback information to setup a primary user (PU)-aware routing technique exploiting the compressive sensing to gain the sparse nature of the PUs occupation \cite{bedewy2015cooperative, kakalou2017cognitive}. The packet size optimization has also introduced a valuable solution for prolonging network lifetime \cite{majumdar2017packet}. Different CR spectrum detectors have been evolved, most known of which: Energy, cyclostationary, and matched filter (MF) detectors \cite{yucek2009survey,sun2013wideband,akyildiz2008survey}. MF was not only limited for CR applications but its use has been extended to blood vessel detection \cite{cabric2004implementation,salahdine2015matched,al2007improved,youssif2007optic}.   
Indeed, CR opens a wide technological gate for many competitive networks to cooperate with for enhancing the day-to-day challenges such as security front and power consumption management. 

In this regard, secured WSNs can dramatically contribute in CR networks (CRNs) as the fusion center (FC) needs to rely on distributed sensor nodes (SNs) to sense the spectrum status. In other words, WSNs suffer from security and power management problems specially in harsh environments. The uncontrolled WSNs security is a double defect: one is data deterioration and second is losing power consumption in transmission of designated infected data. Consequently, security in WSNs-based CR is a prominent issue that has been considered in \cite{ahmad2015survey,salim2016energy,bukhari2016survey,khan2016cognitive,quan2016cooperative,kakalou2017cognitive}. 


One of the eminent techniques utilized in the literature to handle the security problem in WSNs is game theory \cite{abdalzaher2019non,ref_7,abdalzaher2017usingiet,abdalzaher2017using, abdalzaher2016game}. Game theory is a special optimization branch which tackles the interaction among a set of rational intelligent users that aim to enhance their individual gains in an intelligent and adaptive way \cite{ref_5}. To this end, three main dependent challenges affecting a smooth operation for CR are spectrum sensing, mitigating security threats, and the consequent lost energy. One of the security threats in CRNs is spectrum sensing data falsification (SSDF) attack \cite{chen2008toward}. Here, the defender represents the FC. Therefore, the whole communication system in such CRN can be strictly disrupted and the network performance will be steeply degraded. In addition, this attack can disrupt the delivered reports at the FC, and hence, a wrong decision will be taken by the FC about the spectrum status. Moreover, it does not only reduce the number of correctly received reports (packets) at the FC but also enlarge the energy lost. Since the amount of energy to protect the allotted SNs is limited, an intelligent SNs protection mechanism is needed to minimize the amount of  invested power.

In this paper, we propose a game-theoretic approach using Stackelberg game along with MF detector to enhance WSNs-based CR security against the SSDF attack. To the best of our knowledge, no similar work has been presented in the literature which uses game theory along with MF to improve the detection performance of PU, handling the security threat, manage the consequent lost energy, and enhance the correctly delivered reports to be processed at the FC by which the decision is taken.

The main contributions of the paper are three folds:

\begin{itemize}

\item
The Stackelberg game model is designed along with MF detector to optimally detect the infected spectrum status reports, due to the rational external SSDF attack, transmitted from the deployed SNs. To decrease the complexity of employing the MF for spectrum detection, we utilize TDMA protocol to avoid collision between the delivered SNs reports to the FC. Moreover, TDMA is used to support a robust synchronization between the FC and the deployed SNs, which facilitates the detection and protection process.

\item

For adopting a realistic paradigm, the proposed model considers the potential hardware (HW) failure occurred in the deployed SNs and can distinguishes these nodes among the ones influenced by the attack, while attaining improved performance with the presence of the SNs malfunction.      

\item

The proposed model effectiveness is clarified by the achieved probability detection performance in presence of SSDF attack and HW failure. This achievement is denoted by the percentage of protected SNs reports. In other words, the more protected SNs reports are, the more accurate decision is taken by the FC.  Hence, the model can conserve the lost power of reports (packets) retransmission as a result of failure communication attempts due to the SSDF attack impact. The effectiveness of the above proposed approach is verified through simulation evaluation over six different environments, outdoor line-of-sight (OL), outdoor non-line-of-sight (ON), underground line-of-sight (UL), underground non-line-of-sight (UN), indoor line-of-sight (IL), and indoor non-line-of-sight (IN).

\end{itemize}

The rest of the paper is organized as follows. Section \ref{related} presents the related work. In Section \ref{system model}, the system model is discussed. The proposed game formulation is then presented in Section \ref{game formulation}, while the SNs reports delivery is addressed in Section \ref{delivery}. Section \ref{results} shows the obtained results. Finally, the paper conclusion is revealed in Section \ref{conclusion}.

\section{Related Work}\label{related}

This section discusses the related works to enhance the CRNs performance including the security aspect. In the literature context, the security in CR can be the essential metric to guarantee the data privacy and to achieve robust communication system, specifically against the intelligent attacks manipulations such as eavesdropper, primary user emulation (PUE), jamming, byzantine, and SSDF attacks \cite{soltanmohammadi2012decentralized}. As an eminent adaptive and intelligent tool to address the above security aspects, game theory can be utilized in CRN \cite{fang2018three, alsaba2018zero, vu2019repeated,yazdi2019countermeasure,jia2019game,abrardo2014decision}. 

The eavesdropper and jamming attacks effect on CRNs, which recognizes the CRN data privacy, were effectively studied using Stackelberg game, zero-sum game, repeated game \cite{fang2018three, alsaba2018zero, vu2019repeated}. The problem of a full-duplex active eavesdropper, which represents full duplex mode of a jammer and a classical eavesdropper, has been mitigated by a three stage Stackelberg game \cite{fang2018three}. Similarly, the potential eavesdropper in CRN has been confronted by a zero-sum game \cite{alsaba2018zero}. A repeated game was also utilized to build a multiple-channels communications secuity mechanism for SUs against the random eavesdroppers. The efforts of game theory was also extended to confront the PUE attack in CRN to minimize the miss detection using a nonzero-sum game model \cite{yazdi2019countermeasure}. Game theory was exploited to formulate the anti-jamming channel selection problem as an anti-jamming dynamic game in CRN \cite{jia2019game}. In \cite{abrardo2014decision}, a zero-sum game was developed to model the corrupted nodes due to the Byzantine attack, which negatively influences the the CRN routing reputation. The harmful impact is caused by the SSDF attack on the FC decision. Nevertheless, only a few works were presented to resolve this devastating issue in the CRNs. 

In \cite{abdalzaher2017usingiet}, a Stackelberg game was utilized in order to mitigate the SSDF attack to detect the corrupted nodes reports in WSNs-based CR. Meanwhile, in \cite{wang2014trust}, the problem of fake inspections sent by malicious SUs in CRNs due to the SSDF was studied using static game approach to establish a statistical trust model. In \cite{abdalzaher2017usingiet,wang2014trust}, the energy detection method has been employed with game theory to detect the malicious nodes behavior due to the SSDF attack. The energy detection is the simplest method of detection in CR but has crucial weaknesses, e.g., it can be deceived by the intellectual attack manipulations and cannot reach the optimal detection. To solve this issue, 
more intelligent game model needs to be designed along with a robust detection scheme against the SSDF attack.

Unlike the previous related works, this paper presents an effective game model designed with the MF detector and TDMA based protocol to achieve an optimal solution against the security issue due to the SSDF attack under existence of a possibility of HW failure. In addition, the model is utilized to manage the consequent lost energy over the six different environments, where Tmote Sky based SN model \cite{kurt2017packet} is used to assume a realistic CRN system.

\section{System model}\label{system model}
\begin{figure}[t!]
\begin{center}
\includegraphics[width =\columnwidth]{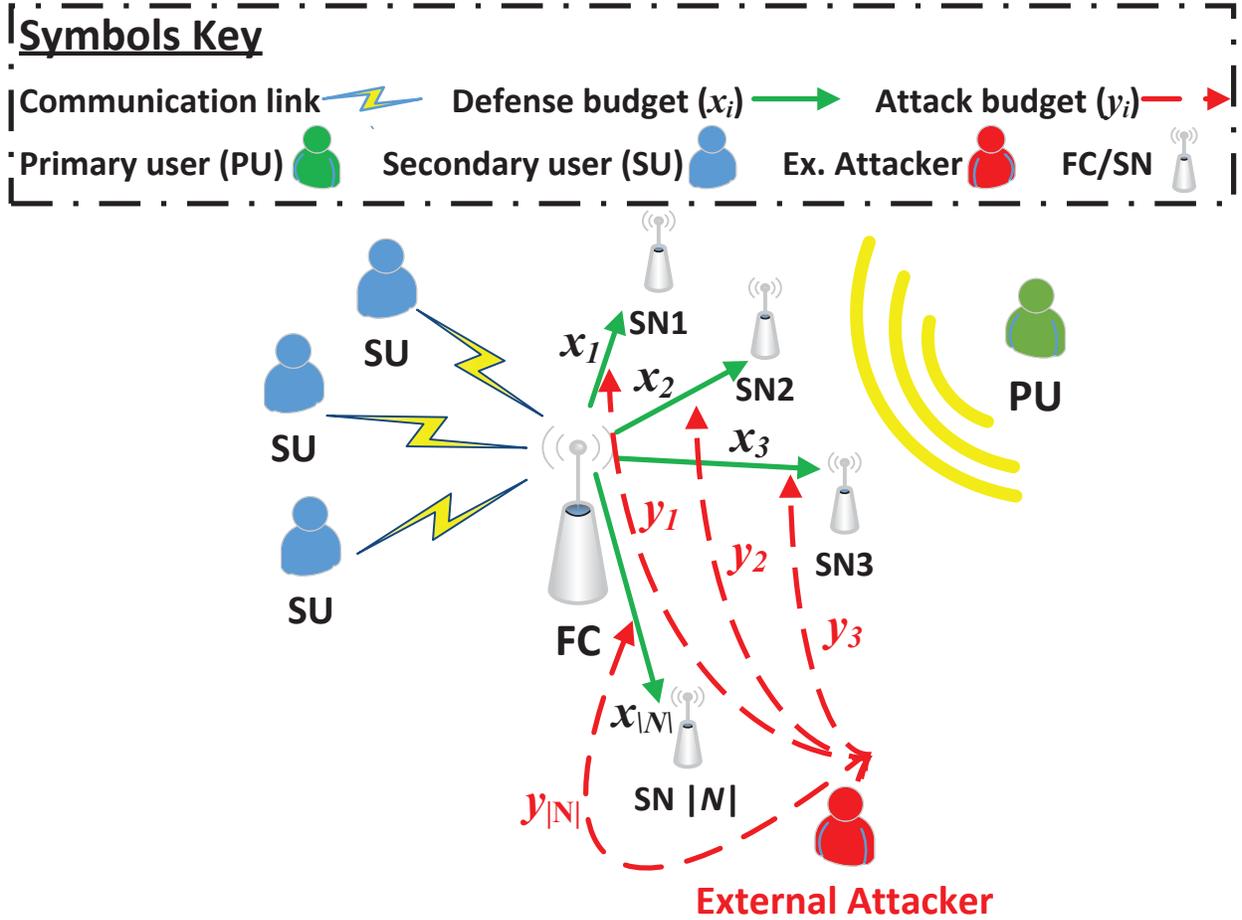}
\caption{WSNs-based CR basic structure.}
\label{sys_arch}
\end{center}
\end{figure}

The proposed model is considered in Fig. \ref{sys_arch}. Notations and variables related to the proposed model and its mathematical representation are summarized in Table \ref{list}. The FC relies on $\mathcal{|N|}$ SNs that are utilized to identify the spectrum holes. The SNs send the sensed spectrum as reports to the FC as shown in Fig. \ref{sys_arch}. Using this sequence, we aim to provide a collision free of spectrum access to the secondary users (SUs) based on the delivered protected SNs reports to the FC against the SSDF attack. We assume that the SSDF attack aims to devastate the communications between the SNs and the FC in WSNs-based CR over statics AWGN channel, i.e., all SNs are symmetrically distributed around the FC, where the distances between FC and all SNs are the same.

Figure \ref{sys_arch} indicates the system structure of the proposed model for protecting the $i$-th SN report ($i \in \{ 1,2, \cdots, \mathcal{|N|} \}$) from the impact of SSDF attack. The distributed defense budgets ($x_i$)s on the SNs reports by the FC are used to countermeasure the attack budgets ($y_i$)s. Indeed, the defense and attack budgets can represent the amount of power invested in the defense and attack actions, respectively. More particularly, the attack budget ($y_i$) is exploited as interference signal power, noise added to the original signal (report), to corrupt the transmitted reports from each SN to the FC. Therefore, we consider that the SSDF deteriorates the SNR by injecting noise power. The reports transmission from SNs to FC is organized based on TDMA to prevent the collision problem.

On the other hand, the FC defense budgets represent the desired transmission power of each SN to neglect the SSDF attack impact and hence the performance is examined at the equilibrium point. The allotted defense budgets are represented by the green arrows to countermeasure the SSDF attack effect. The red icon is the external attacker that applies the attack budgets represented by the green arrows to disrupt the communication between the SNs and FC. To this point, in the next section, we propose a Stackelberg game to model this interaction between the FC and the external attacker, which represents the leader/follower nature of the problem; the FC acts as the leader and the attacker acts as a follower. It should be noted that the attacker actions will be affected by the actions taken by the FC. Then, the defense budget can be utilized as an internal self-defense mechanism to protect the delivered reports from the potential corruption due to the rational SSDF attack manipulations.

\begin{figure}[t!]
\begin{center}
\includegraphics[width =\columnwidth]{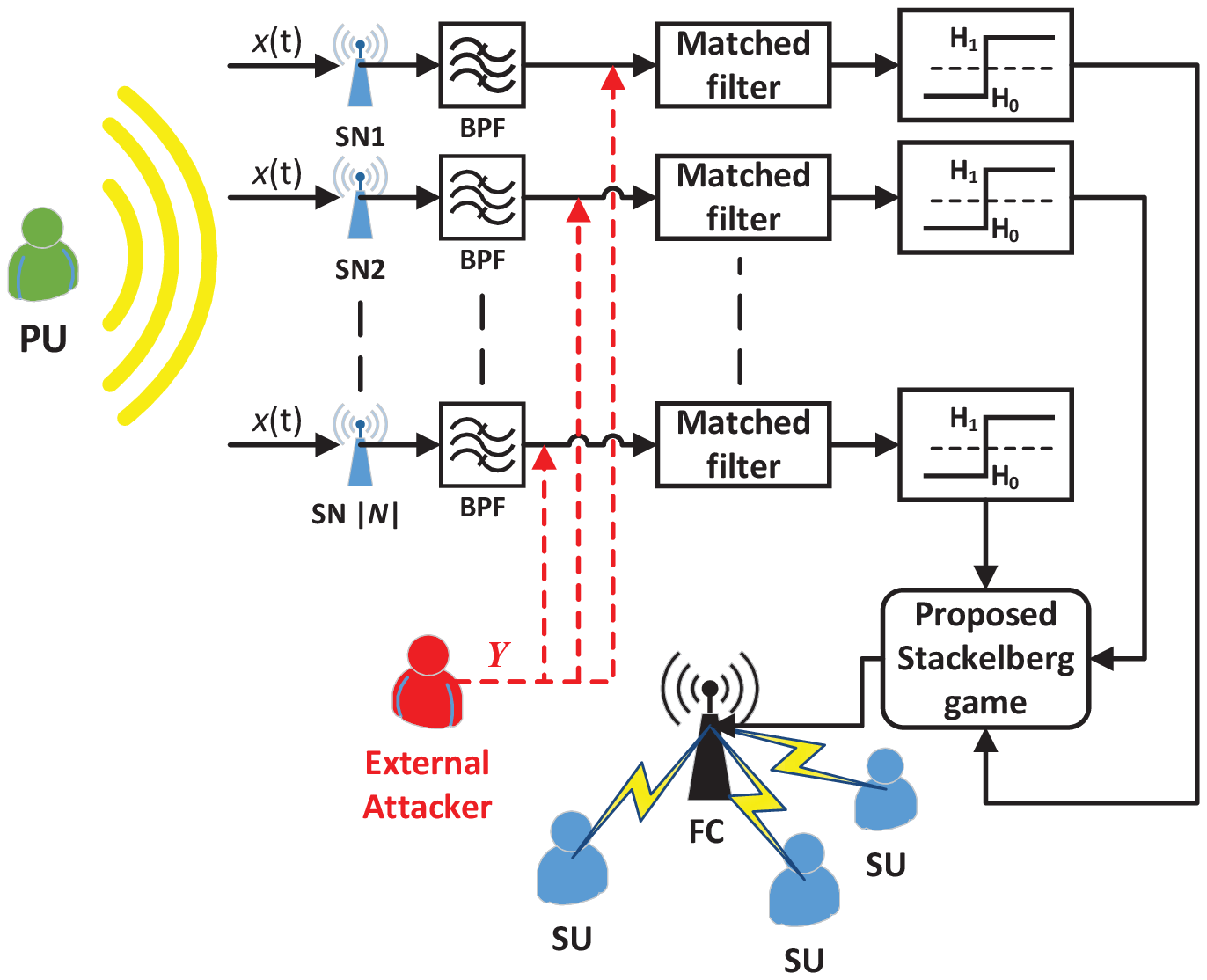}
\caption{Game process along with MF detector with the presence of the attack.}
\label{MF}
\end{center}
\end{figure}

Figure \ref{MF} shows the system structure along with the MF. It is clearly noted that the process starts by the deployed SNs to sense the spectrum followed by BPF. The external attacker intervenes aiming at disrupting the designated delivered report to FC at this stage. The FC is in charge of taking a decision about being the PU is present or absent. This decision is based on the combined SNs reports passed the thresholding level. 

The final decision is represented by a binary hypothesis testing problem. The two hypotheses in Eq. (\ref{hybo}) are given in presence of noise $(H_{0})$ or noise plus signal $(H_{1})$ taking into account the interference noise power added by the attack where we assume that all SNs have the same noise floor.
\begin{equation}\label{hybo}
\begin{split}
H_{0}:y_{i}(n)  &= CG_i . s(n), \text{PU absent} \\
H_{1}:y_{i}(n) &= CG_i . s(n)+\text{w}_{i}(n), \text{PU present}
\end{split}
\end{equation}
where $y_i(n)$ is the SN received
signal, $CG$ is the complex channel gain of the sensing channel, $s(n)$ is the PU signal, $\text{w}(n)$ is the additive white Gaussian noise (AWGN) with zero mean and variance $\sigma_i^2$,$n=1,2,...,N_{sm}$, $N_{sm}$ is the sample number, $i=1,2,...,\mathcal{|N|}$.

The optimal filter that projects the received signal in the direction of the pilot $x_p$ \cite{weidling2005framework} can be written as follows:
\begin{equation}\label{}
\begin{split}
c_{i}  = \sum^{N_{sm}}_{n=1}  y_i(n)x^*_p(n).
\end{split}
\end{equation}

Using the Neyman-Pearson (NP) detector, it is well known that differentiating between the two hypothesis $(H_{1}/H_{0})$ is based on the test statistics ($c_{i}$) \cite{salahdine2015matched} using a threshold $\lambda_{i}$ as follows.
\begin{equation}\label{eq3.4}
\begin{aligned}
c_{i}\frac{\overset{H_{1}}{>}}{\underset{H_{0}}{
<}}\lambda_i,\; i=1,2,...,\mathcal{|N|}.
\end{aligned}
\end{equation}

Accordingly, the probability of detection $P^{i}_{d}$ and the probability of false alarm $P^{i}_{f}$ at $i$-th SN are already influenced by the attack interference power, and can be given by
\begin{equation}\label{pd}
\textit{P}^{\textit{i}}_{\textit{d}}=Pr(c_{i}>\lambda_{i}|H_{1})=Q\left(\frac{\lambda_{i}-E
}{\sqrt[]{E\sigma_i^2}}\right).
\end{equation} 
\begin{equation}\label{pf}
\textit{P}^{\textit{i}}_{\textit{f}}=Pr(c_{i}>\lambda_{i}|H_{0})=Q\left(\frac{\lambda_{i}
}{\sqrt[]{E\sigma_i^2}}\right),
\end{equation} 
where $Q(\cdot)$ is a function that provides the tail probability of the standard normal distribution at $i$-th SN. The threshold level of every $i$-th report depends on the power level captured by each. $E$ is the PU signal energy. Hereafter, we omit superscript $i$ for simplicity of notation. The sensing threshold ($\lambda$) is a function of PU signal energy and noise variance giving by
\begin{equation}
\lambda = Q^{-1}(\textit{P}_{\textit{d}}) \sqrt[]{E\sigma_i^2}
\end{equation}


Indeed, the FC confesses the received report based on matching filter followed by a thresholding level to determine that whether this report withstands the external attack interference power ($y_i$) or not. This interference power aim is to deviate the SNR ($\gamma$) from its interference-free value as

\begin{equation}\label{snr}
\gamma_i = \frac{G (S_i)}{P_n + y_i}, i \in {1,2,\cdots, |\mathcal{N}|}
\end{equation}
where $G$ denotes channel gain. $S_i$ and $P_n$ represent transmission power of the $i$-th SN report and noise power, respectively.

Thus, the obtained $P_d$ or $P_f$ actually do not reflect the real spectrum observation. Consequently, we propose a Stackelberg game that aims at observing the attack interference power, detecting the infected reports, and hence enhancing the detection performance. Then, SNR is given by

\begin{equation}\label{eq_eq_1}
\gamma_i = \frac{ G (S_i+x_i)}{nP+y_i}, i \in {1,2,\cdots, |\mathcal{N}|}
\end{equation}

Therefore, the used ($x_i$) at the equilibrium point can neglect ($y_i$) in Eq. (\ref{snr}) as

\begin{equation}\label{eq_eq_2}
\tilde{\gamma}_i = \frac{G (S_i+ x^*_i)}{ P_n +y^*_i} \implies \frac{G (S_i)} {n}
\end{equation}
where $x^*_i$ and $y^*_i$ are the optimal defense and attack budgets at the equilibrium point for the $i$-th report.


\section{Game Formulation}\label{game formulation}

In this paper, the MF detector is used in WSNs-based CR along with Stackelberg game features. To this end, in this section, we propose a Stackelberg game model, where the competition is between the FC and the external attacker. The attacker strategy depends on the defender (FC) strategy and vice versa. The model is developed to confront the SSDF attack effect. More concretely, this approach concentrates on detecting the corrupted reports delivered to the FC in WSNs-based CR. Those corrupted reports aim to deceive the FC leading to a wrong decision taken by the FC.  

In fact, the attacker selects to disrupt a group of SNs reports that have an energy protection budget below a specific thresholding level $\xi$. This mechanism follows the process shown in Fig. \ref{MF}. It is worth mentioning that, the effect on the performance, the relationship of the probability of detection $P_{d}$ and false alarm $P_{f}$, is evaluated at the equilibrium point.

The proposed model target is to protect the used SNs reports in the WSNs-based CR that are exploited to pinpoint whether the spectrum is idle or busy. The game model structure is based on a sequence of actions expressed by two players: a leader/defender ($L$), where ($L$) initiates the game, while the other player follower/attacker ($F$) replies with an action that yields to its optimal utility given the action $L$ \cite{ref_5}.

The utilities/payoffs functions of $L$ and $F$ are denoted by the $U_{L}(A_{L},A_{F})$ and $U_{F}(A_{L},A_{F})$, respectively, where $A_{L}=x_{i}$ and $A_{F}=y_{i}$ are the individual actions taken by $L$ and $F$, respectively. The leader starts with a hypothetical action $A_{L}^{0}$ as the game assumes that $L$ has knowledge about $U_{F}$, which is utilized to withstand the attacker deceptions.
Consequently, $L$ can determine the optimal action $A_{F}^{*}(A_{L}^{0})$ of $F$ Eq. (\ref{eq3.22}) supposing that $F$ is rational player, given that $A_{L}^{0}$ has been exerted. Therefore, the optimal action of $L$ can be given by Eq. (\ref{eq3.23}) {\cite{abdalzaher2017usingiet}}.

\begin{equation}\label{eq3.22}
A_{F}^{*}(A_{L}^{0})= \operatorname*{arg\,max}_{A_{F} \in \bar{\textit{\textbf{A}}}_F} U_{F}(A_{L}^{0},A_{F}). 
\end{equation}

\begin{equation}\label{eq3.23}
A_{L}^{*}= \operatorname*{arg\,max}_{A_{L}^{0} \in \bar{\textit{\textbf{A}}}_L} U_{L}(A_{L}^{0},\operatorname*{arg\,max}_{A_{F} \in \bar{\textit{\textbf{A}}}_F} U_{F}(A_{L}^{0},A_{F})).
\end{equation}
Four definitions are now in order:

\begin{itemize}

\item

\textbf{Definition 1}: The FC defense strategy is represented by $\text{DS}(X)$, where $X$ is the total defense energy budget, which is indicated by the amount of electric power invested in the defense set of actions ($\bar{\textbf{\textit{A}}}_d$). The action performed to protect $i$-th SN  is $A_x = x_{i},i \in \{1,2, \cdots, \mathcal{|N|}\}$, and the defender (FC) set of actions $\bar{\textit{\textbf{A}}}_d$ satisfy constraint $X \geq \sum_{i=1}^{\mathcal{|N|}}x_{i}$. More particularly, this defense budget is used to protect the SNs reports against the attack manipulations represented by injecting noise power to the observed signal to disrupt the delivered report to FC.  

\item 
\textbf{Definition 2}: The attack strategy is indicated as $\text{AS}(Y)$, where $Y$ is the total attack energy budget which is represented by the amount of electric power invested in the attack set of actions ($\bar{\textbf{\textit{A}}}_a$). The action performed to attack $i$-th SN is $A_a = y_{i} = [0,1], i \in \{1,2, \cdots, \mathcal{|N|}\}$, and the attacker set of actions $(\bar{\textit{\textbf{A}}}_a)$ satisfy constraint  $\sum_{i=1}^{N}y_{i}\leq Y$. More concretely, this defense budget is utilized to manipulate the attacker and mitigate its impact by countervailing the added noise power.  

\item

\textbf{Definition 3}: The utilities (payoffs) functions of FC and attacker are represented by $U_{d}(A_{d}) = U_{d}(x_i)$ and $U_{a}(A_{a}) = U_{a}(y_i)$, respectively.

\item

\textbf{Definition 4}: We can achieve subgame Nash equilibrium (NE) at ($A_d^* = x^*_i \; \forall i$) and ($A_a^* = y^*_i \; \forall i$) as shown in (\textbf{Appendix A}). NE can be defined as a set of strategies such that none of the participants in the game can improve their payoff, given the strategies of the other participants. In other words, NE can be explicitly defined as no player can improve its utility by changing its action/strategy, if the other players conserve their current actions/strategies \cite{ref_5}.

\end{itemize}

The leader seeks to enhance its payoff using an appropriate $A_{L}^{*}$ given that both $L$ and $F$ actions have been done in sequence. Consequently, $L$ can accurately handle the only available action of rational $F$ among $(\bar{\textit{\textbf{A}}}_F)$ as 

\begin{equation}\label{eq3.24}
A_{F}^{*}= \operatorname*{arg\,max}_{A_{F}\in \bar{\textit{\textbf{A}}}_F} U_{F}(A_{L}^{*},A_{F}).
\end{equation}

Therefore, subgame NE can be attained using Eqs. (\ref{eq3.22}, \ref{eq3.23}, \ref{eq3.24}). It is clear that, in the examined WSNs-based CR, the game played by the defender (FC) representing the $L$ and the attacker acting as $F$. The defender aims to protect the SNs delivered reports, while the attacker attempts to disrupt these reports. The interactions of $L$ and $F$ are executed over the network lifetime. 

The proposed Stackelberg game model is explicitly organized using \textbf{Algorithm 1} for detecting and protecting the WSNs-based CR from the delivered malicious SNs reports to the FC. Firstly, the FC distributes the defense budgets ($x_i$)'s to the deployed SNs for protecting them against the SSDF attack effect. On the other hand, the attacker allots the corresponding attack budgets ($y_i$)'s attempting to turning the maximum number of SNs reports into malicious. In fact, the attacker investigates every previous action done by the FC to protect the SNs reports. When the game starts, the utility function of the received reports ($U_i$) is investigated. This function denotes the difference between the defense and attack budgets on the $i$-th SN report as given by
\begin{equation}\label{ui}
U_i = x_i - y_i.
\end{equation}

\begin{algorithm}[t]
\caption{Proposed Stackelberg game Algorithm for SSDF attack. 
}
\SetAlgoLined
 \textbf{Distribute} $x_i$\;
 \textbf{Input} $A_d = x_i$, $A_a = y_i$ \;
 \While{NE does not exist (more SNs reports are still infected)}{
  Compute $U_{i}$ by Eq. (\ref{ui}), $\forall i \in |\mathcal{N}|$\;
  Determine the malicious nodes/HW failure \;

	\If{$U_i < 0$}
   {\eIf {This is $1^{st}$ time $U_i < 0$}{
   This $i$-th node $\rightarrow$ HW failure lis\;}
   {This $i$-th node $\rightarrow$ SA malicious list\;}}

 \If{ $U_i > 0$}{
 {This $i$-th node is still benevolent\;}
     \eIf {$U_i > \xi$}{ 
Add this $i$-th node to SL\;}{
Add this $i$-th node to WL\;}
}
Sort SL and WL in ascending order based on the resulting $U_i$\;
Top of heap of SS will be the strongest node (SSR)\;
 \If{ $SL(2) > SL(1)$}
 {Resort SS in descending till finishing all WS\;} 
Defender resorts ($x_i$)'s distribution based on Eqs. (\ref{def_strong},\ref{def_weak})\;

Attacker resorts ($y_i$)'s distribution based on Eqs. (\ref{att_strong},\ref{att_weak})\;
Check ($A_{d},A_a$) $\to$ ($A^*_{d},A^*_a)\; \forall i \in \mathcal{N}$\;
Compute $U_i$ using Eq. (\ref{ui})\;
}
 \textbf{Output} NE exist $\gets$ ($(A^*_{d}, A^*_a), \forall i$)
\end{algorithm}

If $U_i$ is less than zero, this $i$-th report is checked whether it is a result of a HW failure or by the SSDF effect. Otherwise, this $i$-th SN report is still benevolent. More particularly, to realize the proposed model, we assume a percentage of the SNs can suffer from a HW failure during the communication. Then, if the resulting $U_i$ is noncontinuous negative, this  is due a HW failure for the corresponding SN.  Conversely, if the value of this $U_i$ is continuous negative, this $i$-th SN report is considered infected by the SSDF attack and will be excluded from the FC consideration.   

The defense mechanism of the FC is to protect the maximum number of delivered reports leading to accurate spectrum status decision. Consequently, the defense scheme relies on the benevolent received reports about the spectrum. Indeed, the thresholding level ($\xi$), which is application sensitivity dependent, is utilized to distinguish the strongly and weakly protected SNs reports that are sorted in strong list (SL) and weak list (WL), respectively; given that the attacker budget is limited. The result of Eq. (\ref{ui}) determines the intelligent attack strategy in every new round. In this strategy, the attacker concentrates on the weak list (WL) of the SNs reports that have been protected with low defense budgets in the previous round. The attacker sorts the WL in a descending order based on Eq. (\ref{ui}) result. It means that the utility function of those reports satisfies the following constraint
\begin{equation}\label{utility}
U_i = (x_i - y_i) < \xi.
\end{equation}

Based on the normal feature of the Stackelberg game, the leader knows the attack strategy and thus, it will apply the same concept as the follower behaves to counter the attacker effect. Accordingly, F exerts a very little extra attack budget $\alpha$ deducted from its attack budget applied on the
strongest SN report (SSR) in the previous iteration to be added to the weak SN report (WR) as

\begin{equation}\label{att_strong}
y_{w}(\text{SSR})|_{new} =y_{w}(\text{SSR})|_{old}-(U_{r}+\alpha), 
\end{equation}
\begin{equation}\label{att_weak}
y_{r}(\text{WR})|_{new}=y_{r}(\text{WR})|_{old}+(U_{r}+\alpha).
\end{equation} 
where $w$ is an index of a strongly protected SN, and $r$ is an index for a weakly protected SN. This redistribution strategy of attack budget is performed to make sure that $U_{r}$ in the next iteration will be lower than zero; which  compromising the weakly protected SN report $r$. 

In contrary, the FC manipulates the attacker by adding $\xi$ budget value to each weak SN report that has $(U_{i}<\xi)$ after subtracting enough ($\xi$)'s from the strongly protected SN report (SSR) to satisfy the requirements of weak SN report in the next iteration as given in the following equations:
\begin{equation}\label{def_strong}
x_{w}(\text{SSR})|_{new}=x_{w}(\text{SSR})|_{old}-(\xi).
\end{equation}
\begin{equation}\label{def_weak}
x_{r}(\text{WR})|_{new}=x_{r}(\text{WR})|_{old}+(\xi).
\end{equation}

\begin{figure}[t]
\begin{center}
\includegraphics[width =\columnwidth]{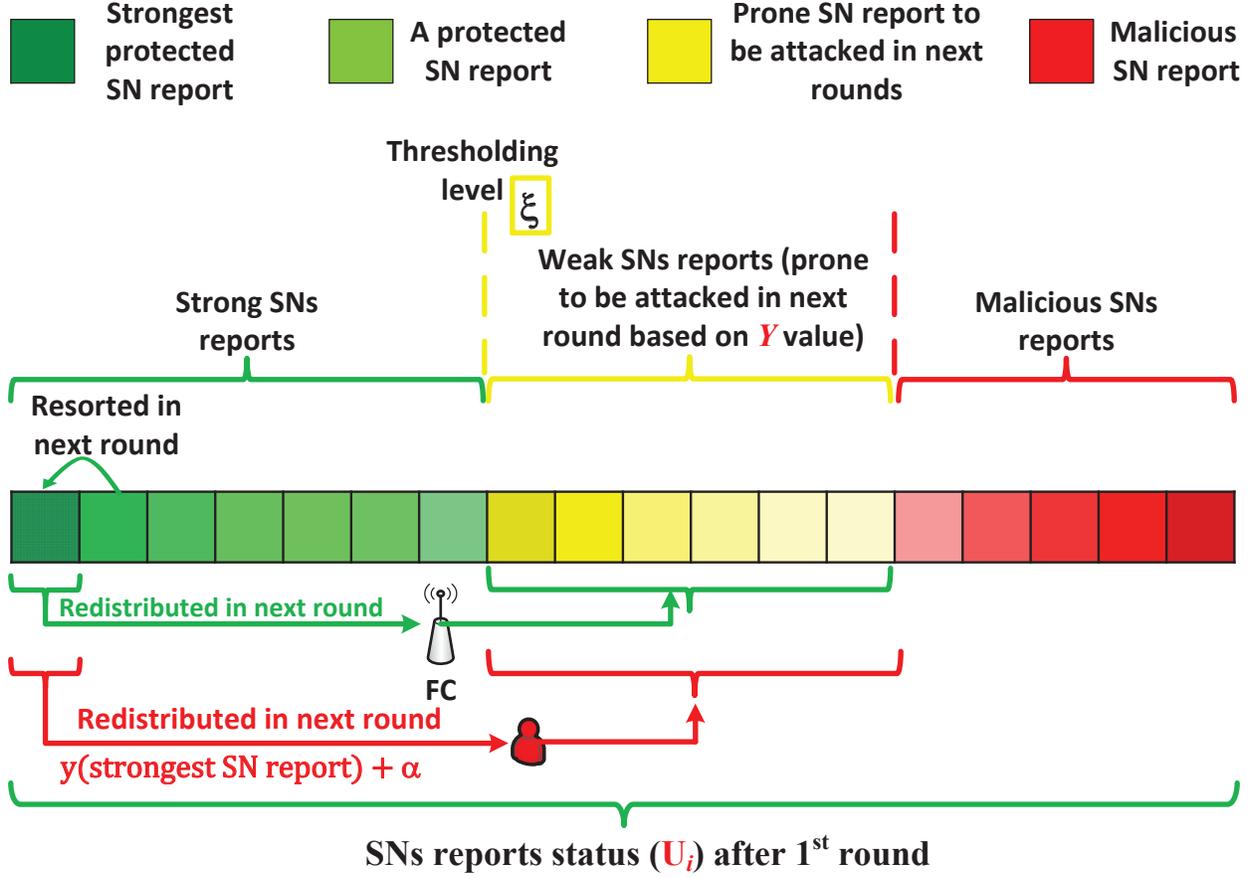}
\caption{Graphical process of the mathematical induction.}
\label{math_induc}
\end{center}
\end{figure}

Figure \ref{math_induc} graphically shows the mathematical induction used for the WL and SL of SNs reports leading to the NE at which no more negative effect added by the attacker. Obviously, the green cells indicates the SL of protected SNs reports in which SSR is the top of heap (darkest green cell). After first iteration, the FC deducts a $\xi$ value from the previously applied defense budget of the SSR, i.e., $x(SSR)-\xi$, to be re-allotted on the WL (prone to be attacked in the next rounds). Consequently, the SL will be rearranged every subsequent iteration. The thresholding level ($\xi$) is the border line between SL and WL. Then, it is followed by the yellow cells representing the list of weakly protected SNs reports. Conversely, the attacker will add an $\alpha$  attack budget to the previously applied attack budget of the SSR, i.e., $y(SSR)+\alpha$, to be redistributed on the WL. Finally, the red cells denote the malicious SNs reports.    

Finally, this model makes sure that the system is reliable, given that the attacker has a limited budget. Consequently, when we deal with pure actions/strategies where the competitive players (FC and external attacker) choose deterministic actions, the NE is defined here as follows:

\begin{equation}\label{NE}
U_i(A_{d}^*,A_{a}^*) \geq U_i(A_d,A_{a}^*) \: \forall A \in \textbf{\textit{A}}, i \in \mathcal{N}.
\end{equation} 

\section{Reports Delivery Formulation}\label{delivery}

\begin{figure}[t]
\begin{center}
\includegraphics[width =\columnwidth]{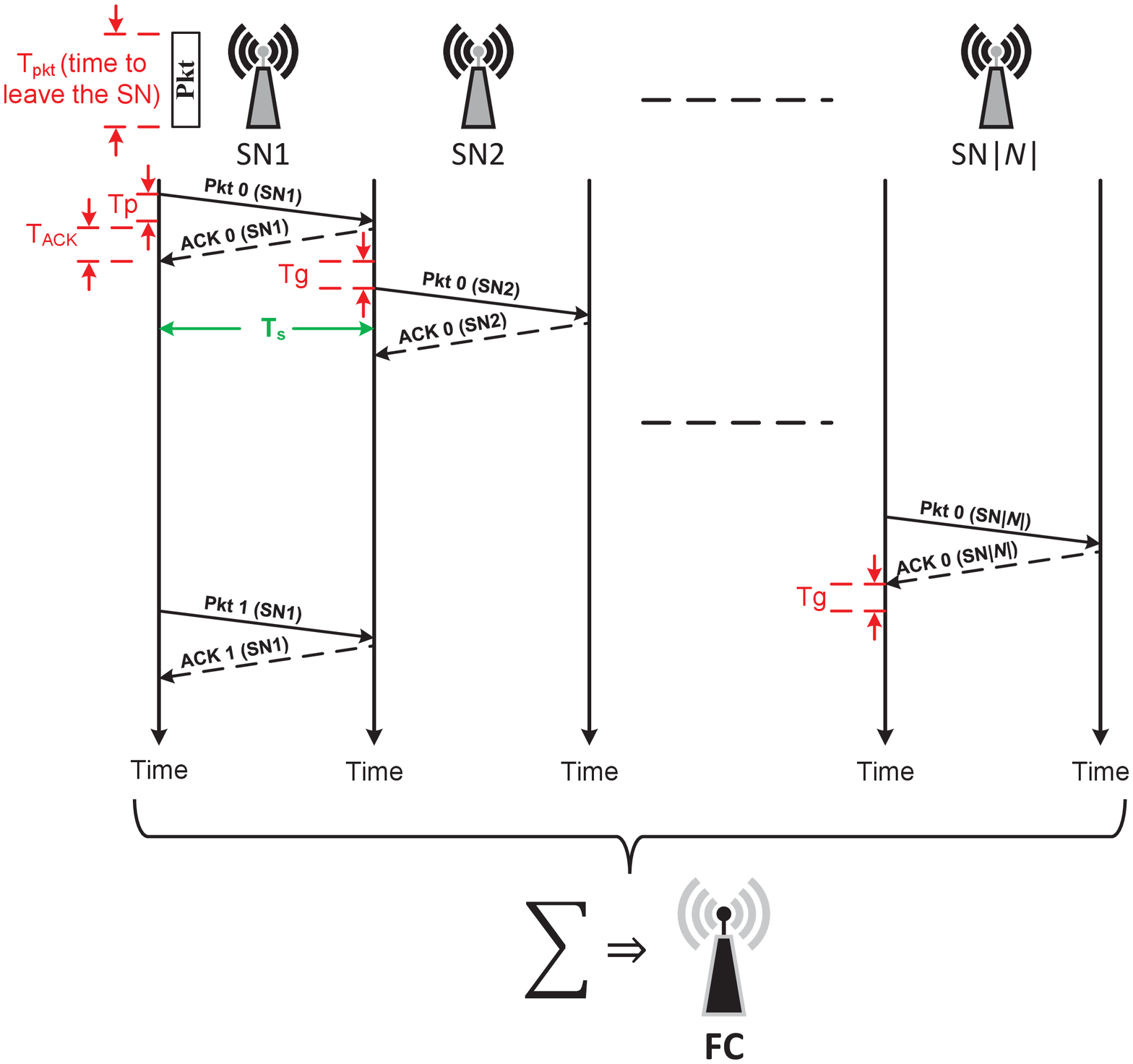}
\caption{Time diagram based on TDMA technique.}
\label{time}
\end{center}
\end{figure}

This section explains the considered system model analysis and the reports (packets) delivery formulation. We assume that TDMA is used for data packet transmission to prevent the possibility of collision. TDMA time slots allocations are assigned by the FC as depicted in Fig. \ref{time}. We consider a star topology between the FC and the set of SNs observers, $\mathcal{N}$, where the number of SNs in the network is given by the cardinality of the set $\mathcal{N}$ which is represented by  $\left| \mathcal{N} \right|$. 

In fact, the communication protocol between the SNs and the FC is based on transmitting a data packet from $i$-th SN to the FC and then the FC replies by an acknowledgment (ACK) packet to confirm receiving the data packet. This communication sequence is called a successful handshake. 

\begin{table}[h!]
\centering
\caption{Notations of parameters and variables.}
\label{list}
\begin{tabular}{|c | l |}
\hline 
 Symbol & Description \\
  \hline \hline
   $\mathcal{N}$ & The set of sensor nodes\\
    \hline 
    $\mathcal{Z}$ & The set of all nodes excluding the FC \\
    \hline
    $X$ & \begin{tabular}[c]{@{}l@{}}The total defense budget \end{tabular} \\
    \hline
        $x_i$ & \begin{tabular}[c]{@{}l@{}}Applied defense budget for $i$-th report \end{tabular} \\
    \hline
    $Y$ & \begin{tabular}[c]{@{}l@{}} The total attack budget\end{tabular} \\
    \hline
    $y_i$ & \begin{tabular}[c]{@{}l@{}}Applied attack budget for $i$-th report \end{tabular} \\
    \hline
    $T_s$ & Time slot duration  \\
    \hline
    $T_g$ & Guard time between two consecutive time slots  \\
    \hline
    $T_t$ & Transmission time duration  \\
    \hline
    $T_p$ & Propagation time duration \\
    \hline
    $T_b$ & Total busy time duration \\
    \hline
    $pkt$ & Data packet length  \\
    \hline
    $ACK$ & Acknowledgment packet length  \\
    \hline
     $D$ & Payload length contained in a data packet  \\
    \hline
     $H$ & Header length contained in a data packet  \\
    \hline
    $PL_{iFC}$ & \begin{tabular}[c]{@{}l@{}} The path loss of the link between the every $i$-th SN and FC\end{tabular} \\
    \hline
    $PL_{0}$ & \begin{tabular}[c]{@{}l@{}} The free space path loss at the reference distance of the \\antenna far field \end{tabular}\\
    \hline
    $d_{iFC}$ & The distance between the communicating nodes $i$ and FC  \\
    \hline
    $d_{0}$ & The reference distance of the antenna far field   \\
    \hline
    $n$ & The path loss exponent  \\
    \hline
    $\sigma$ & The standard deviation of the shadow fading  \\
    \hline
    $P^A_{r,iFC}$ & \begin{tabular}[c]{@{}l@{}}  The received signal power at the antenna of FC from \\transmitting node $i$ \end{tabular}   \\\hline
$P^A_{t}$ & The transmission signal power at the transmit antenna  
      \\
      \hline
      $P_{ct}$ & The transmission power consumption\\
    \hline
    $\gamma$ & The signal-to-noise ratio\\
    \hline
    $BER$ & The bit error rate\\
    \hline
    $P_{n}$ & The receiver noise power\\
    \hline
    $L$ & The number of bytes per packet \\
    \hline
    $Pr_{G}$ & The process gain\\
    \hline
    $P^S$ & The probability of successful packet reception\\
    \hline
     $P^F$ & The probability of failure packet reception\\
    \hline
     $P^{SHS}$ & The probability of successful handshake\\
    \hline
     $P^{FHS}$ & The probability of failure handshake\\
    \hline
     $R_e$ & \begin{tabular}[c]{@{}l@{}} The probability of re-transmission to achieve successful\\ handshake \end{tabular}  \\
    \hline
     $E_{dt}$ & The energy dissipation for transmitting a packet\\
    \hline
     $E_{dpp}$ & The energy dissipation for processing a packet\\
    \hline
     $E_t^{HS}$ & The total energy dissipation\\
    \hline
    $T_A$ & The data acquisition time duration\\
\hline
    $m$ & A transmit energy level of the data packet\\
    \hline
    $u$ & A transmit energy level of the ACK packet\\
    \hline
     $P_{off}$ & The power consumption during sleep mode\\
    \hline
    $B$ & The total energy of the sensor node battery  \\
    \hline
    $E_{DA}$ & The energy dissipation due to data acuasition per round \\
    \hline
    $P_{DA}$ & The power dissipation due to data acuasition  \\
    \hline
    $T_{DA}$ & The data acquisition time \\
    \hline
    $P_{cr}$ & The power consumption due to data reception  \\
    \hline
\end{tabular}
\end{table}


The adopted path loss between every $i$-th SN and the FC $(i, FC)$ is given by 
\begin{equation}\label{eq3}
PL_{iFC} [\texttt{dB}] = PL_{0} [\texttt{dB}] + 10nlog_{10} \frac{d_{iF}}{d_0} + \sigma [\texttt{dB}], 
\end{equation}
where $PL_{0}$ is the free space path loss at the reference distance $d_0$ of the antenna far field, $n$ denotes the path loss exponent, $d_{iFC}$ is the distance between the transmitting $i$-th SN and the FC, and $\sigma$ represents the standard deviation in dB of the shadow fading. 

The received antenna signal power at the FC from the $i$-th transmitting SN with power level $m$ can be given by 
\begin{equation}\label{eq4}
P^{A}_{r,iFC}(m)  [\texttt{dBm}]=  P^{A}_{t}(m) [\texttt{dBm}] - PL_{iFC}  [\texttt{dB}],
\end{equation}
where $P^{A}_{t}(m)$ is the transmission antenna signal power at power level $m$. Table \ref{table_1} illustrates the transmission power consumption with the eight available power levels ($P_{ct}(m)$) and the antenna transmission power $P^A_{t}$ using Tmote Sky node \cite{corporaton2006tmote}.

\begin{table}[t]
\centering
\caption{Consumed transmission power $P_{ct}$ and antenna output power $P^A_{t}$ with every power level $m$ \cite{corporaton2006tmote}.}
\label{table_1}
\begin{tabular}{|l|l|c|l|l|c|}
\hline 
  $P_{ct}$ (mW)& $P^A_t$ (dBm) & $m$ & $P_{ct}$ (mW)& $P^A_t$ (dBm) & $m$ \\
  \hline \hline
    25.5 & -25 &  3 & 41.7  &  -5 & 19 \\
    \hline 
    29.7   & -15 & 7 & 45.6  &  -3 & 23 \\
    \hline 
     33.6   & -10 & 11  &  49.5  &   -1& 27 \\ \hline
     
     37.5 & -7 & 15 & 52.2 &   0 &   31   \\
    \hline 
    
\end{tabular}
\end{table}

The obtained SNR for the signal power transmitted from $i$-th node to the FC is given by 
\begin{equation}\label{eq5}
\gamma_{iFC}(m)  [\texttt{dB}]=  P^{A}_{r,iFC}(m) [\texttt{dBm}] - P_n  [\texttt{dBm}],
\end{equation}
where $P_n $ is the receiver noise power. Table \ref{table_2} depicts the path loss parameters for the available six environment measurements presented in \cite{gungor2010opportunities}.
\begin{table}[ht!]
\centering
\caption{Path loss parameters for the six environments \cite{gungor2010opportunities}.}
\label{table_2}
\begin{tabular}{|m{4cm} | c | c | c |}
\hline 
 Environment & $n$ & $\sigma (dB)$ & $Pn (dBm)$ \\
  \hline \hline
   Outdoor-LOS (OL) & 2.42  & 3.12 &-93  \\
    \hline 
    Outdoor-NLOS (ON)   & 3.51  & 2.95 &-93  \\
    \hline 
     Underground-LOS (UL)   & 1.45  & 2.45 & -92 \\ \hline
 Underground-NLOS (UN) & 3.15  & 3.19 & -92  \\
    \hline 
  Indoor-LOS (IL)  &  1.64  & 3.29 & -88   \\
    \hline 
    Indoor-NLOS (IN)  &  2.38  & 2.25 & -88 \\ \hline
\end{tabular}
\end{table}
In fact, the BER based on the used SN (Tmote Sky) is given by
\begin{equation}\label{eq6}
BER = Q \bigg( \sqrt[]{\frac{2E_b}{N_0}} \bigg),
\end{equation}
where $\frac{E_b}{N_0} = \gamma_{iFC} (m)Pr_G$, $Pr_G$ denotes the process gain which is the ratio between the chip rate and the bit rate of the spread spectrum system \cite{corporaton2006tmote}.


Thus, a successful packet reception probability of an uncoded $L$-Byte packet transmitted at power level-$m$ among the communicating terminals $(i, FC)$ is given by
\begin{equation}\label{eq8}
p_{iFC}^{S}(m,L) = \Bigg ( 1-Q\bigg( \sqrt[]{16\gamma_{iFC}(m)} \bigg) \Bigg)^{8L}.
\end{equation}
On the other hand, the failure probability of receiving a packet is given by
\begin{equation}\label{eq9}
p_{iFC}^{F}(m,L) = 1 - p_{iFC}^{S}(m,L). 
\end{equation}

The probability of a successful handshake denotes the probability that a data packet is  successfully transmitted from the $i$-th node to the FC ($p_{iFC}^{S}$) with power level ($m$) and a successfully ACK packet is responded by the FC to the $i$-th SN ($p_{FCi}^{S}$) at power level ($u$), which is given by 
\begin{equation}\label{eq10}
p_{iFC}^{SHS}(m,u) =  p_{iFC}^{S}(m,pkt) \times  p_{FCi}^{S}(u,ACK).	 
\end{equation}
Consequently, the probability of failure handshake is given by 
\begin{equation}\label{eq11}
p_{iFC}^{FHS}(m,u) = 1 - p_{iFC}^{SHS}(m,u).
\end{equation}
Clearly, the data packets should be re-transmitted ($Re_{iFC}(m,u)$) times to ensure that the data packets are successfully transmitted. The number of re-transmissions is given by
\begin{equation}\label{eq12}
Re_{iFC}(m,u) = \frac{1}{p_{iFC}^{SHS}(m,u) }. 
\end{equation}
Energy dissipation for transmitting a ($pkt$)-Byte packet from the $i$-th SN to the FC with power level ($m$) is given by
\begin{equation}\label{eq13}
E_{dt} (m,pkt)= P_{ct}(m) T_{pkt},
\end{equation}

Indeed, when a SN accomplishes the packet transmission, it stays in the receive mode during the time slot. Therefore, this situation should be taken into consideration at calculating the total energy dissipation as

\begin{equation}\label{eq14}
E_{t}^{HS} (m,pkt)= E_{dt}(m,pkt) + P_{cr}(T_{s} - T_{pkt}),
\end{equation}
where $P_{cr}$ represents the power consumption of the data reception during the rest of the time slot, $T_s$ is the slot time, and $T_{pkt}$ represents the actual packet transmission time. Furthermore, the energy consumed for packet processing and for re-transmission until attaining successful handshake is given by 

\begin{equation}\label{eq15}
E_{dt} (m,u)= E_{dpp}  + Re_{ij}(m,u)E_{t}^{HS}(m,pkt),  
\end{equation}
where $E_{dpp}$ is the energy dissipation for packet processing, which is computed only once when successful handshake exists. On the other hand, the successful handshake energy dissipation at the receiver side, $E_{r}^{SHS} (u,ACK)$,  consists of the energy consumed for receiving the data packet and the energy consumed for transmitting ACK to the transmitter. Thus, $E_{r}^{SHS} (u,ACK)$ is given by 

\begin{equation}\label{eq16}
\begin{split}
E_{r}^{SHS} (u,ACK)  =  P_{cr}(T_{s} - T_{ACK})  + E_{t}^{ACK}(u,ACK).
\end{split}
\end{equation}
Conversely, the energy dissipation of failed handshake due to the  absence of data packet is given by 

\begin{equation}\label{eq17}
E_{r}^{FHS} = P_{cr}T_{s}.
\end{equation} 
Consequently, the total energy dissipation at the receiver taking into account all data packets re-transmissions and ACK packets re-transmissions is given by 
\begin{equation}\label{eq18}
\begin{split}
E_{r,FCi}^{SHS} (m,u)  & =   E_{dpp} + Re_{iFC}(m,u)[  p_{iFC}^{SHS}(m,u) \times \\ &  
E_{r}^{SHS}(u,ACK) +  p_{iFC}^{S}(m,pkt) \times \\  & p_{FCi}^{F}(u,ACK) E_{r}^{SHS}(u,ACK) \\  & +  p_{iFC}^{F}(m,pkt)E_{r}^{FHS}]. 
\end{split}
\end{equation}
It should be noted that if a SN is not busy during a specific time slot it stays in the sleep mode, wherein the power consumption is $P_{off}$.

\section{Simulation Results}\label{results}

This section presents the simulation results. The simulation parameters are presented in Table \ref{parameters}. We assume Tmode Sky \cite{gungor2010opportunities,kilic2013analysis} as a realistic SN model and employ the parameters from the data sheet in \cite{corporaton2006tmote}. We assumed that all SNs use power level $m=31$. It is worth mentioning that, the number of generated $pkt$s by every SN per round depends on the selected $D$. For instant, if $D$ equals 120 bytes, only one data packet will be generated by every SN. If $D$ equals 60, two data packets will be generated. We use four different total attack budgets $Y \in \{8,10,13,17\}$ for model verification and $Y= \{8,17\}$ for showing the WSNs-based CR performance using Eqs. (\ref{pd},\ref{pf}) (${P}_{{d}}$ vs.  ${P}_{{f}}$)) and (${P}_{{d}}$ vs.  $SNR$)) with the same FC defense budget $X= 20$. In addition, the FC decision is affected by the observed signal level at the SNs over an AWGN channel taking into consideration the shadowing fading. It is assumed that all SNs are located from the FC with the maximum coverage range of the Tmote Sky SN (125 m) to show the effectiveness of our model. Although the maximum distance is used, the model operates efficiently along with MF detector and TDMA over the six environments (OL, ON, UL, UN, IL, and IN).  

\begin{table}[b]
\caption{Simulation parameters.}
\label{parameters}
\centering
\begin{tabular}{|l|l|l|l|}
\hline
Parameter       & Value     & Parameter   & Value                                                                                                \\ \hline \hline
$\mathcal{|N|}$ & 50        & $T_g = T_p$ & 100 $mu$s \cite{schuts2009modelling}                                                \\ \hline
$P_{cr}$        & 69 mW     & $T_s$       & 4.78 ms                                                                                              \\ \hline
$P_{DA}$        & 11.4 mW   & Data rate   & 250 Kb/s                                                                                             \\ \hline
$T_{DA}$        & 5 ms      & $T_{pkt}$   & $\frac{pkt}{Data rate}$                                                                              \\ \hline
$E_{DA}$        & $mu$J     & $T_{ACK}$   & $\frac{ACK}{Data rate}$                                                                              \\ \hline
$D$             & 120 bytes & $Pr_G$      & $\frac{2 M chip/s}{Data rate}$ \cite{corporaton2006tmote, lanzisera2007theoretical} \\ \hline
$H$             & 8 bytes   & $P_{off}$   & 3 $mu$W                                                                                              \\ \hline
$ACK$           & 12 bytes  & $E_{dpp}$   & 12.66 $mu$J                                                                                          \\ \hline
$m$             & 31        & $B$         & 15 KJ                                                                                                \\ \hline
\end{tabular}
\end{table}

\begin{figure}[!h]
\begin{center}
\includegraphics[width =\columnwidth]{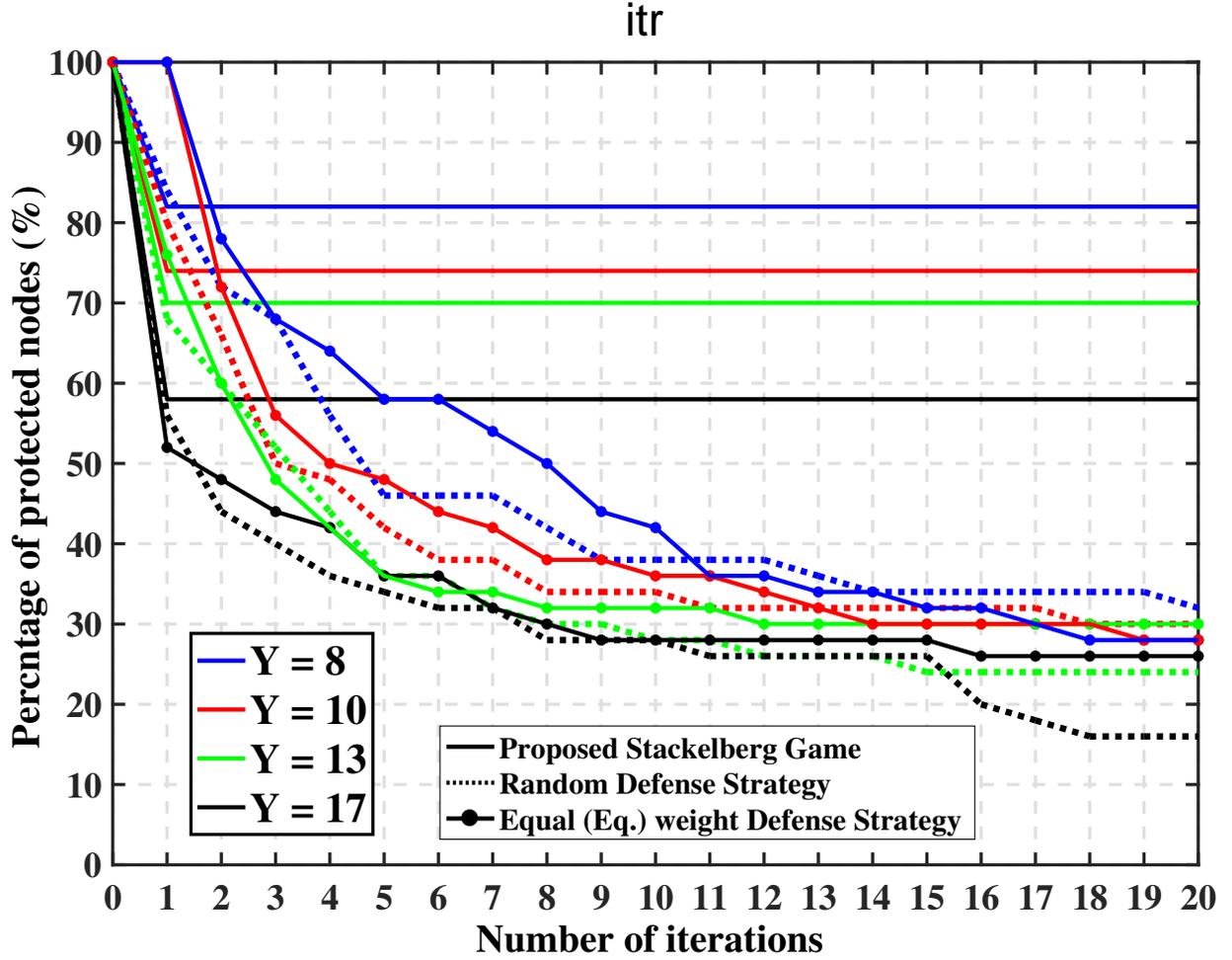}
\caption{The ratio of delivered reports among the total number of delivered reports}\vspace{.1in}
\label{protection}
\end{center}
\end{figure}

Figure \ref{protection} shows the percentage of benevolent SNs reports that withstand the attack manipulations among the total number of reports with the proposed defense mechanism, random defense scheme (the defense power budget is randomly allotted for the SNs) and equal (Eq.) weight defense schemes (the defense budgets are equally weighted distributed for the SNs) within 20 rounds. The percentages of protected SNs reports are about (83$\%$, 74$\%$, 70$\%$, and 58$\%$) due to the applied attack budgets (8, 10, 13, and 17), respectively. Meanwhile, the protected reports percentage steeply degraded while using the random or the equal protection mechanisms. Therefore, it can be estimated from the obtained results that the proposed model achieves efficient protection against the SSDF effect. 

The resulting utility functions of the SNs reports due to the proposed attack-defense strategy are shown in Fig. \ref{utility}. The results indicate the benevolent reports after first round, the reports confronted the attack till the end of simulation, and the ones that turned malicious due to the attack effect, which are colored by yellow bars, green bars, and red bars, respectively. The results prove the success of the proposed model in protecting the majority of reports against the intellectual attack manipulations.   

\begin{figure}[!h]
\begin{center}
\includegraphics[width =\columnwidth, height = 10cm]{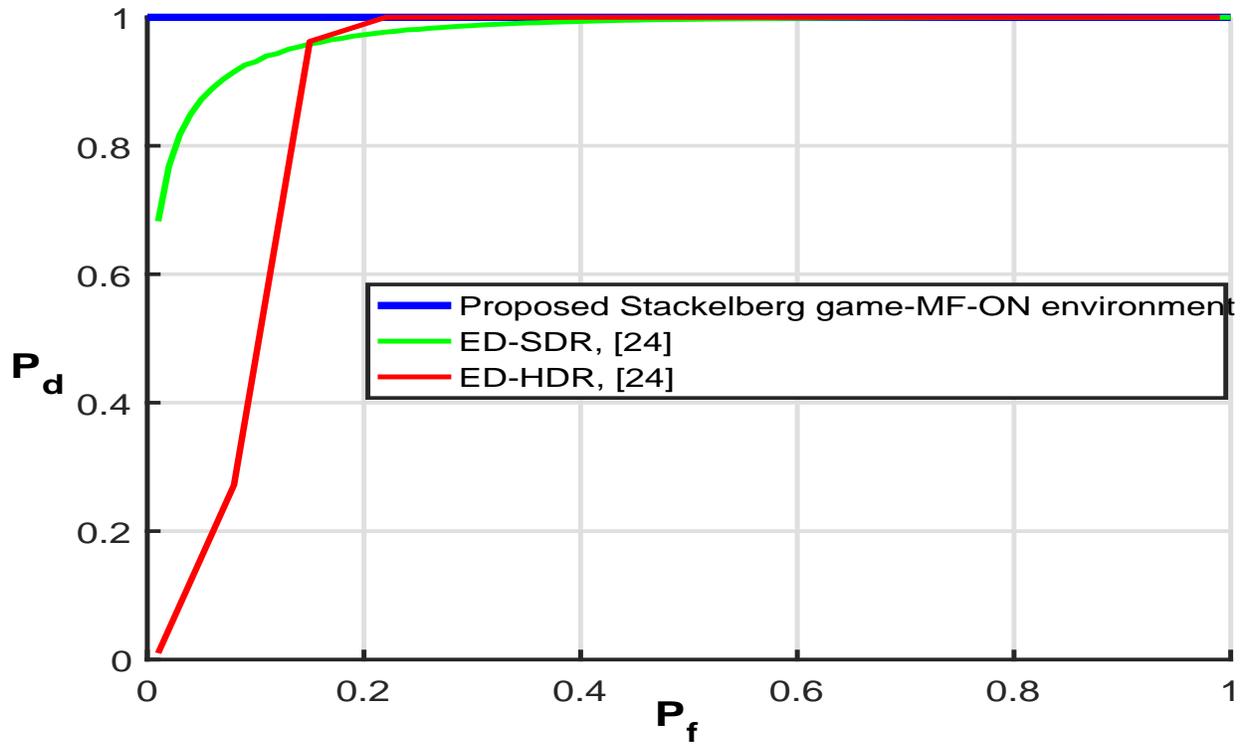}
\caption{$P_d$ vs. $P_f$ comparison of the proposed model and the corresponding in the literature, $Y = 13$.}
\label{pd_pf_model_comp}
\end{center}
\end{figure}

Figure \ref{pd_pf_model_comp} indicates that our model along with the MF outperforms the work in \cite{abdalzaher2017usingiet} in which energy detection (ED) method was used, hard decision rule (HDR) and soft decision rule (SDR). We have shown the results when ($Y = 13$) where in the obtained simulations our model presents better performance even with the other values of attack budget $Y \in \{8,10,17 \}$ over all the six environments. It is worth mentioning that, here, we only show the comparison when ON environment is utilized to indicate the effectiveness of the proposed model with MF among the work done in \cite{abdalzaher2017usingiet} even in the harsh environment. 

\begin{figure}[!h]
\begin{center}
\includegraphics[width =\columnwidth, height = 10cm]{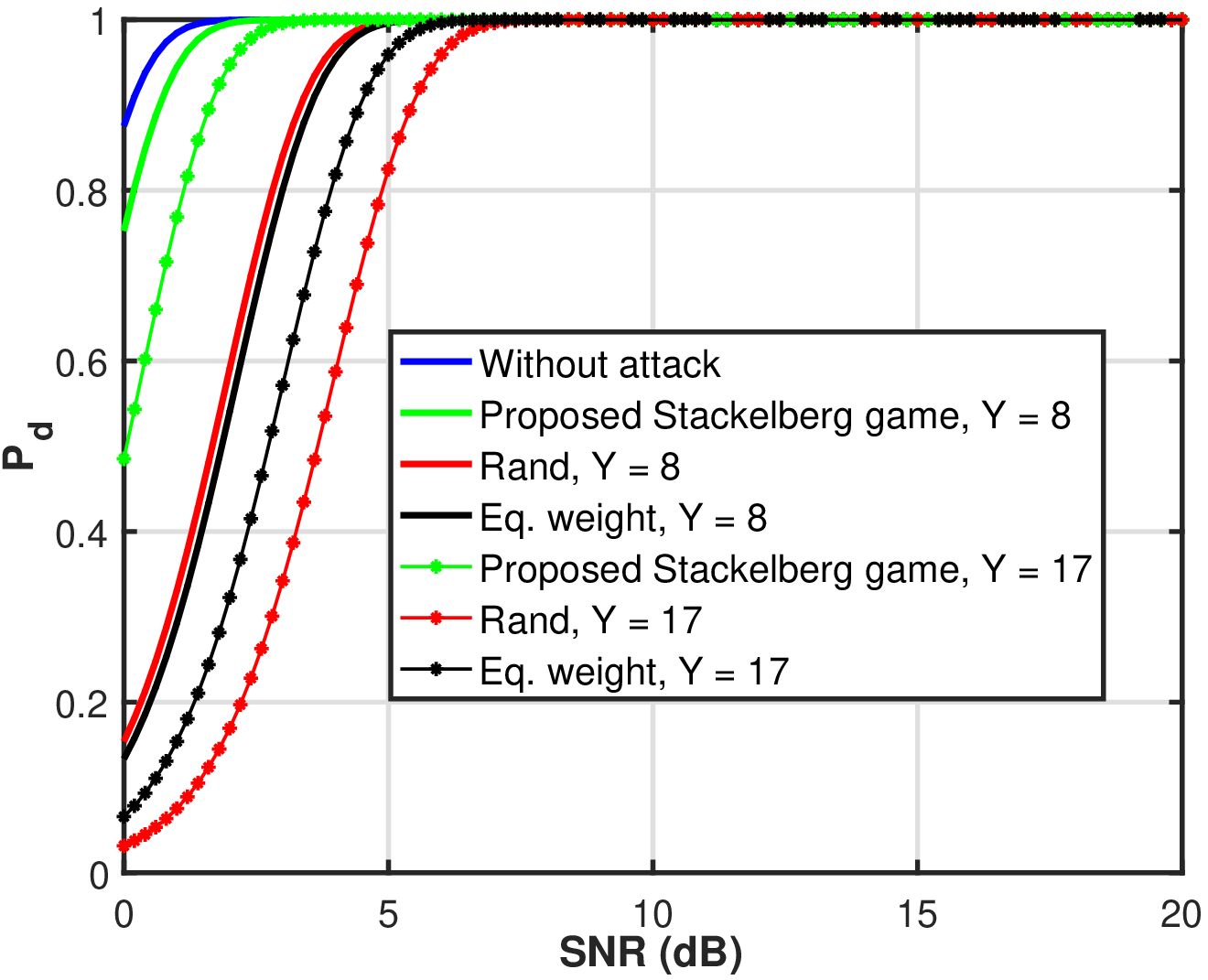}\vspace{-.1in}
\caption{$P_d$ vs. $SNR$ comparison of the proposed model to random and equal weight defense strategies with nonfluctuating PU, $Y=\{8,17\}$.}\vspace{-.1in}
\label{pd_pf_m_8}
\end{center}
\end{figure}

\begin{figure}[!h]
\begin{center}
\includegraphics[width =\columnwidth, height = 10cm]{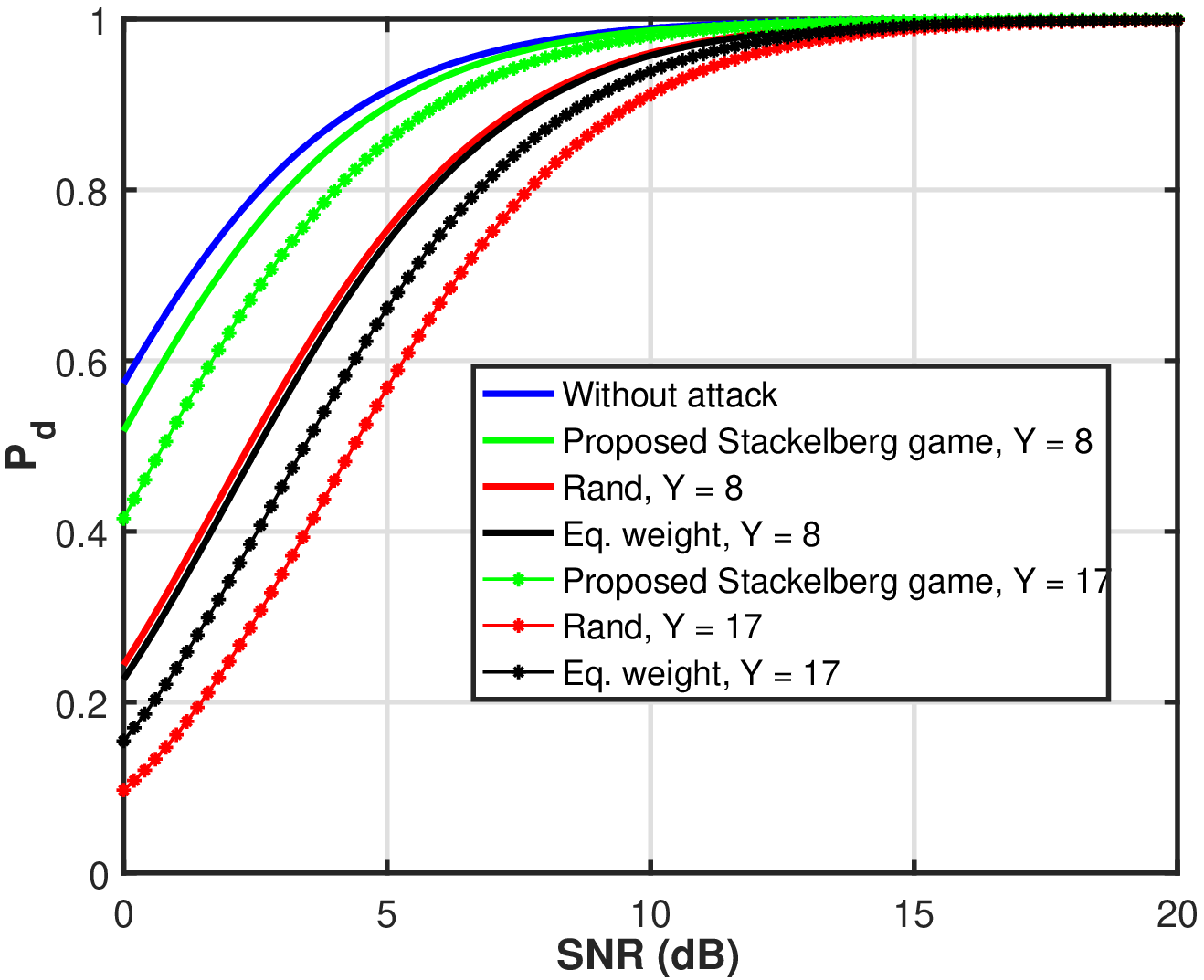}\vspace{-.1in}
\caption{$P_d$ vs. $SNR$ comparison of the proposed model to random and equal weight defense strategies with fluctuating PU, $Y=\{8,17\}$.}
\label{pd_pf_m_13}
\end{center}
\end{figure}

Then, the relationship between $P_d$ and $SNR$ has been extensively studied in two PU cases (nonfluctuating and fluctuating). The obtained results in both cases depict the improved performance whether with fluctuating or non-fluctuating PU as shown in Fig. \ref{pd_pf_m_8} and Fig. \ref{pd_pf_m_13}, respectively. Moreover, we found that changing the environment does not affect the relationship between $P_d$ and $SNR$. Note that, in all simulations done the proposed model presents better performance ($P_d$ vs. $SNR$) as compared to random (rand) and equal (Eq.) weight defense strategies regardless the attack budget value. Therefore, we only show two samples when $Y = \{8, 17\}$, which represent the minimum and maximum used attack budgets, respectively.   

\begin{figure}[!ht]
\begin{center}
\includegraphics[width =\columnwidth, height = 10cm]{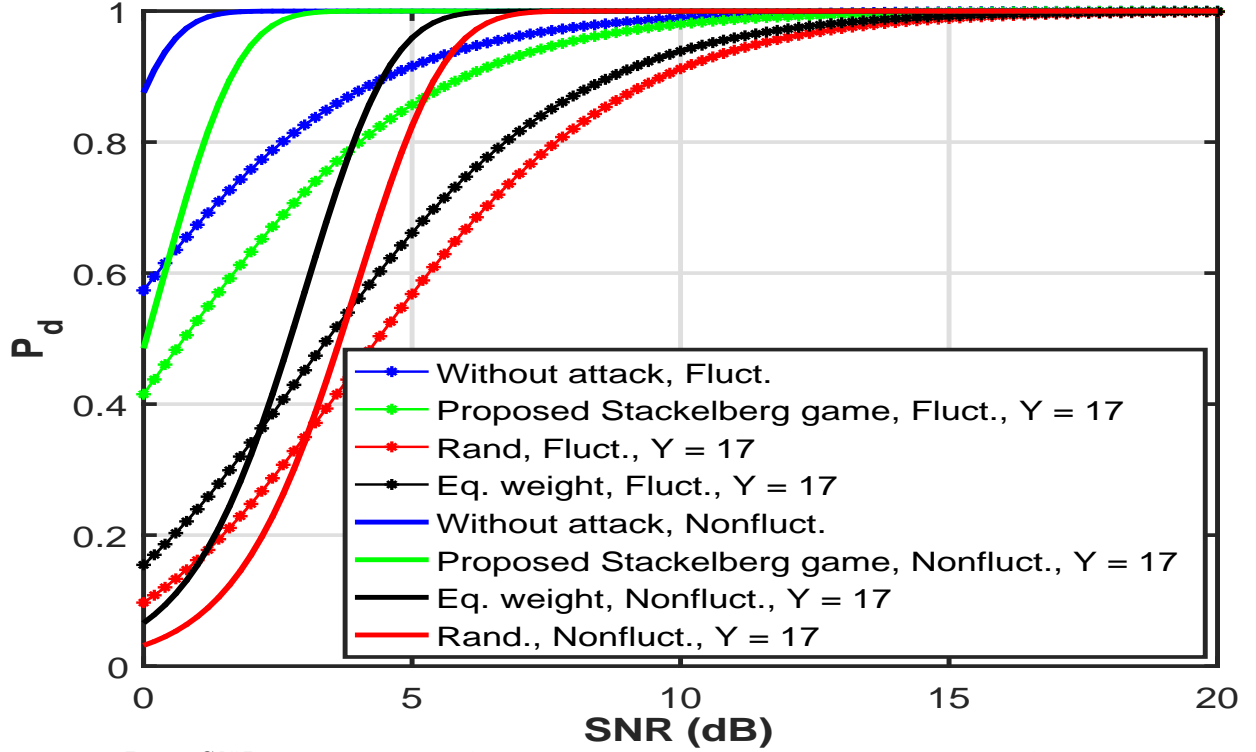}\vspace{-.1in}
\caption{$P_d$ vs. $SNR$ (dB) comparison of the proposed model to random and equal weight defense strategies with NC receiver and fluctuating PU scenarios, $Y = 17$.}
\label{pd_snr_m_8}
\end{center}
\end{figure}

After that, we have selected the maximum attack budget as an indicator of the effectiveness of our model to compare the nonfluctuating (Nonfluct.) and fluctuating (Fluct.) PU obtained curves showing the difference between the proposed model along with MF and without attack attempt applied as shown in Fig. \ref{pd_snr_m_8}. It is clearly shown that the obtained results of the proposed mode is very close to that scenario of without attack. This definitely means that the proposed model with MF can handle the SSDF attack manipulations.

\begin{figure}[!ht]
\begin{center}
\includegraphics[width =\columnwidth, height = 10cm]{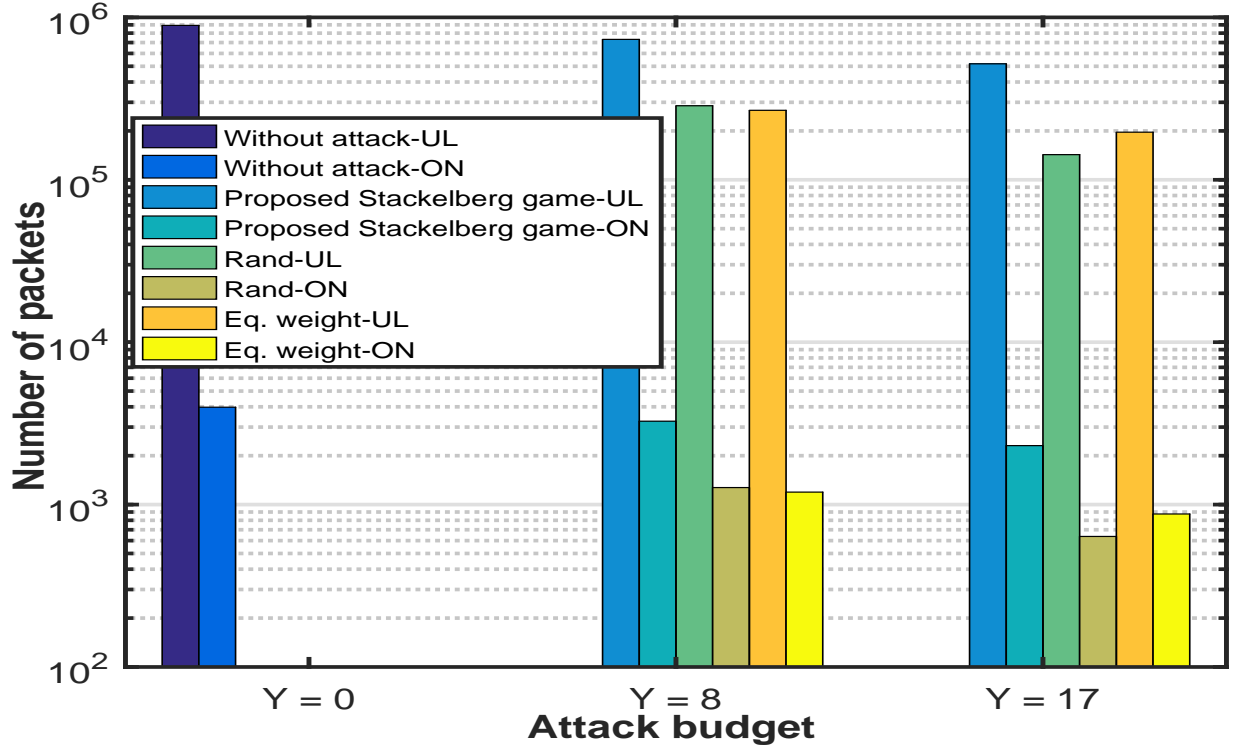}\vspace{-.1in}
\caption{Number of correctly processed packets of the proposed model vs. rand and Eq. weight defense strategies based on (UL) and (ON) environments, $Y = \{8, 17\}$.}
\label{pkt}
\end{center}
\end{figure}

Afterward, the indication of the proposed model on improving the number of correctly packets (protected reports) to be processed by the FC and taking the decision based on is shown in Fig. \ref{pkt}. We have compared our model to the case of no attack exists and both rand and Eq. weight defense strategies over the best environment (UL) and worse environments (ON) when the attack budget is minimum and maximum ($Y=\{8,17\}$). The results ensures the power of the proposed model with MF for confronting the SSDF attack as compared to without attack scenario and definitely with rand and Eq. weight defense strategies. It is clearly shown that the obtained results based on the proposed Stackelberg game model along with MF achieves very close number of packets to be protected to that scenario of no attack whether the attack budget is minimum ($Y=8$) or maximum ($Y=17$).   

\begin{figure}[!ht]
\begin{center}
\includegraphics[width =\columnwidth, height = 10cm]{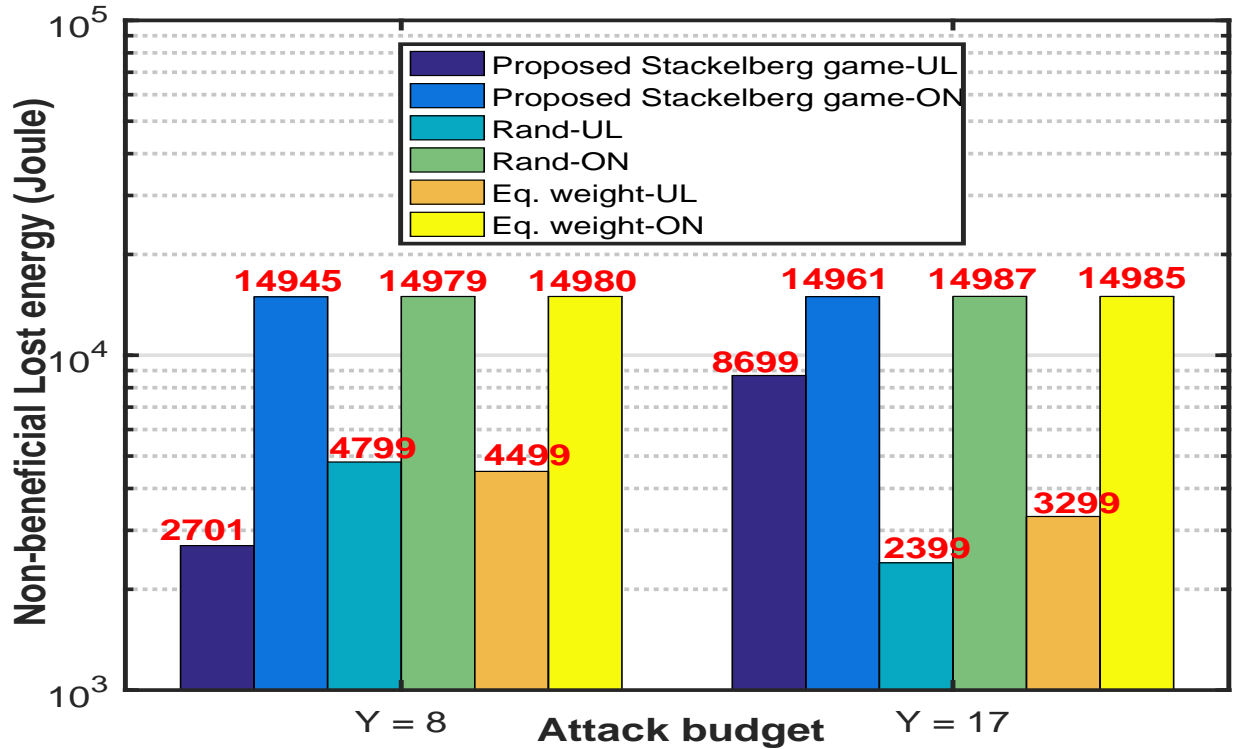}
\caption{The non-beneficial consumed energy using the proposed model vs. rand and Eq. weight defense strategies based on (UL) and (ON) environment, $Y = \{8,17\}$.}\vspace{.1in}
\label{battery}
\end{center}
\end{figure}

Similarly, the lost non-beneficial battery energy of the FC due to consecutive negative feedback (ACK) transmitted to the SNs that its report is infected is shown in Fig. \ref{battery}. We can figure out that the FC battery life time can be maliciously exhausted when the SNs are not effectively protected. It is clear that the harsh environment (ON) adds an extra negative impact plus the attack budget on losing the packets or interfering them which means negative ACK will be sent to those SNs that their packets are affected by the fading problem. Therefore, the results of ON environment get worse as compared to the ones of UL. Moreover, the scenario is more worse when $Y$ increases. Among the extensive obtained simulation results, we show the results when the attack budget equals ($Y = \{ 8, 17\}$) representing the minimum and maximum designated attack budgets, respectively.

\begin{figure*}[!ht]
    \centering 
\begin{subfigure}{0.3\textwidth}
  \includegraphics[width=\linewidth]{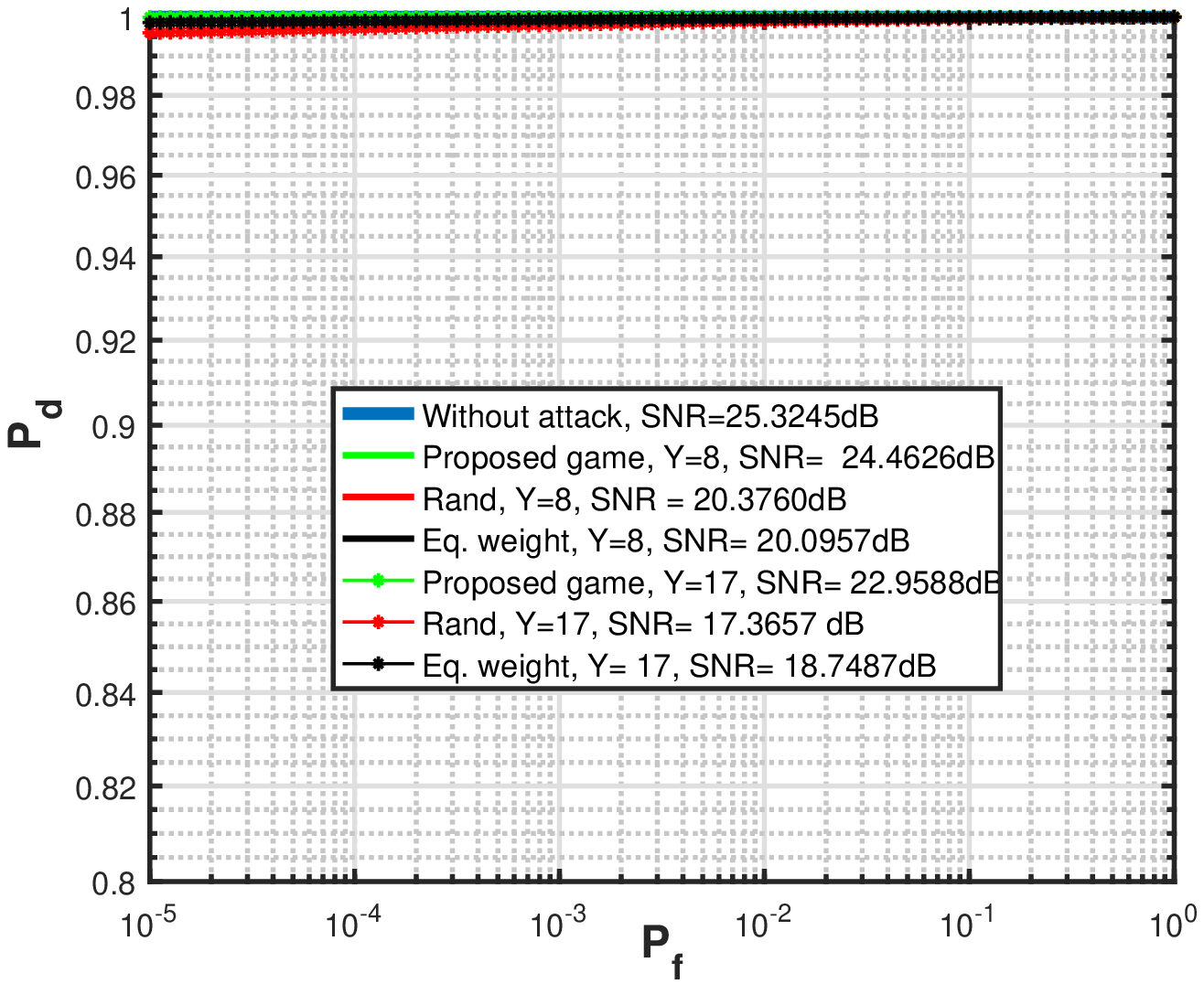} 
  \caption{OL}
  \label{fig:1}
\end{subfigure}\hfil 
\begin{subfigure}{0.3\textwidth}
  \includegraphics[width=\linewidth]{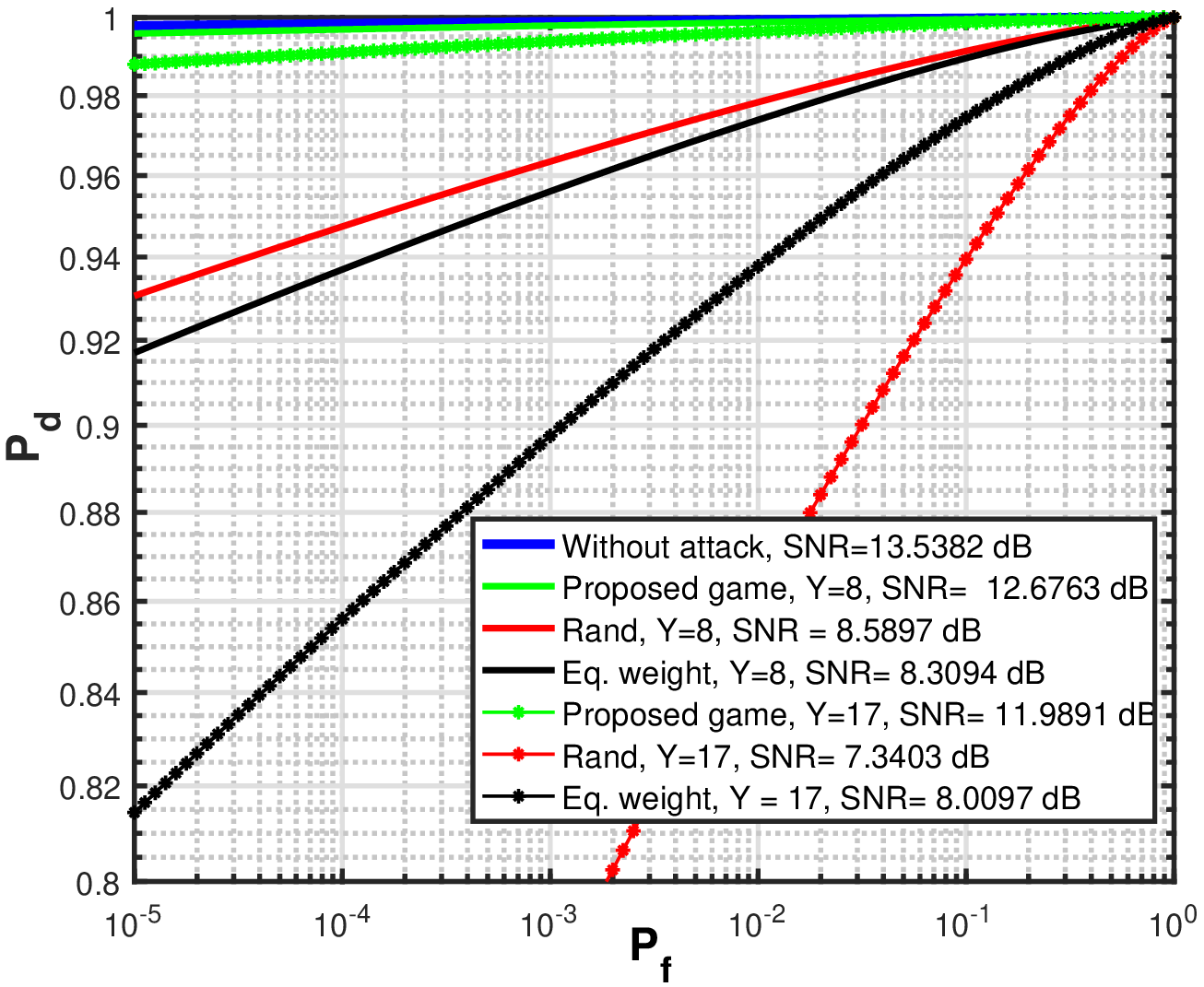} 
  \caption{ON}
  \label{fig:2}
\end{subfigure}\hfil 
\begin{subfigure}{0.3\textwidth}
  \includegraphics[width=\linewidth]{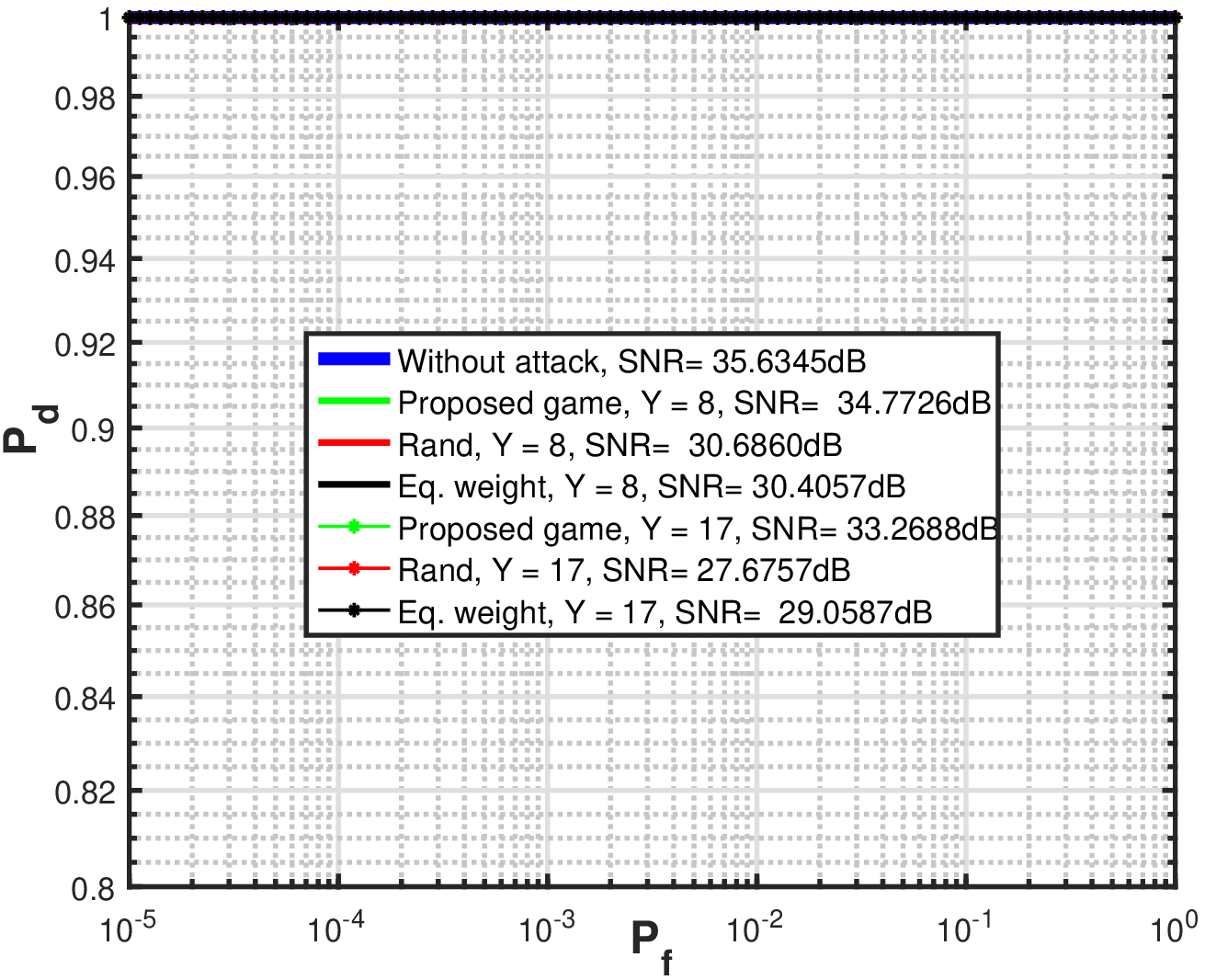} 
  \caption{UL}
  \label{fig:3}
\end{subfigure}

\medskip
\begin{subfigure}{0.3\textwidth}
  \includegraphics[width=\linewidth]{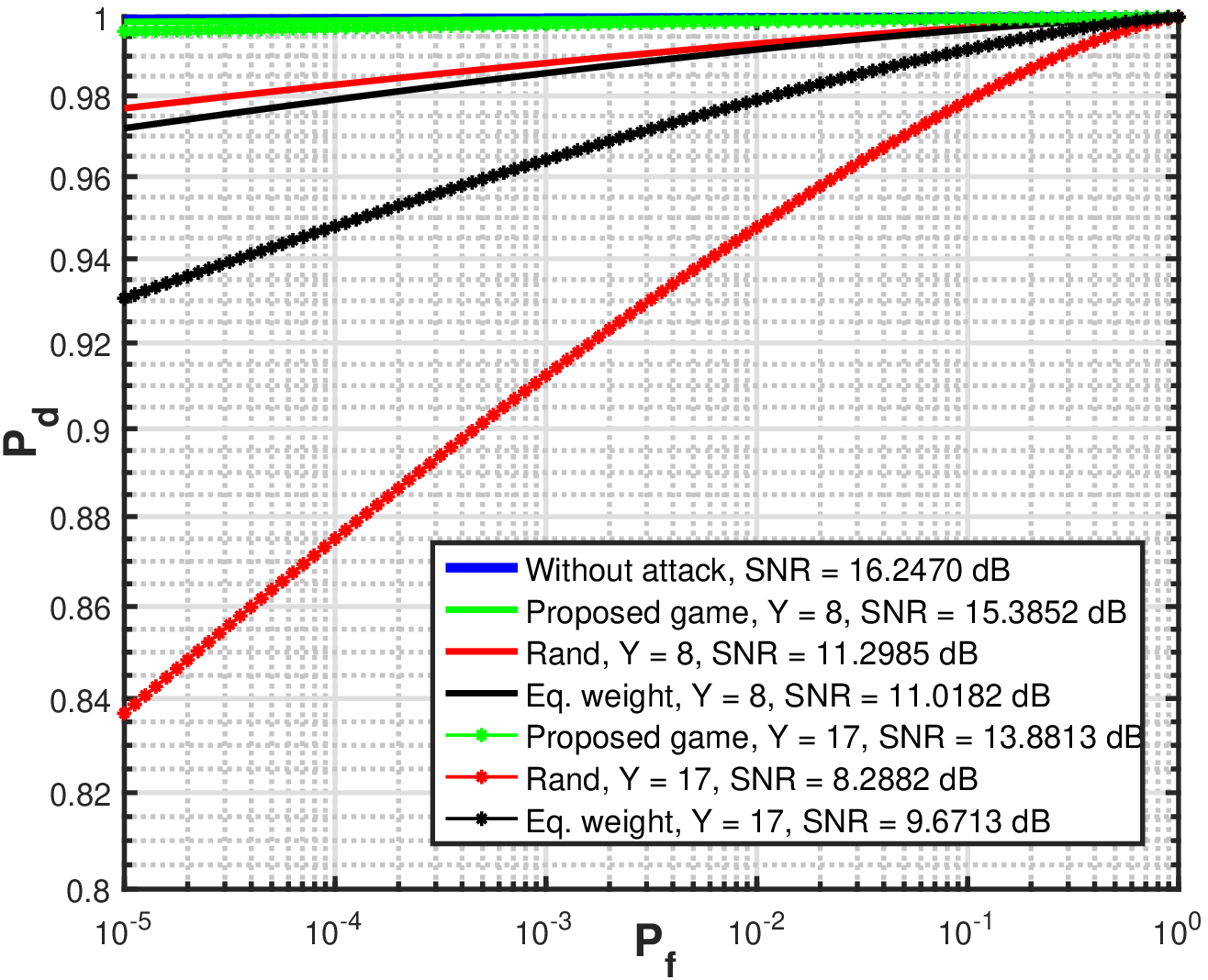} 
  \caption{UN}
  \label{fig:4}
\end{subfigure}\hfil 
\begin{subfigure}{0.3\textwidth}
  \includegraphics[width=\linewidth]{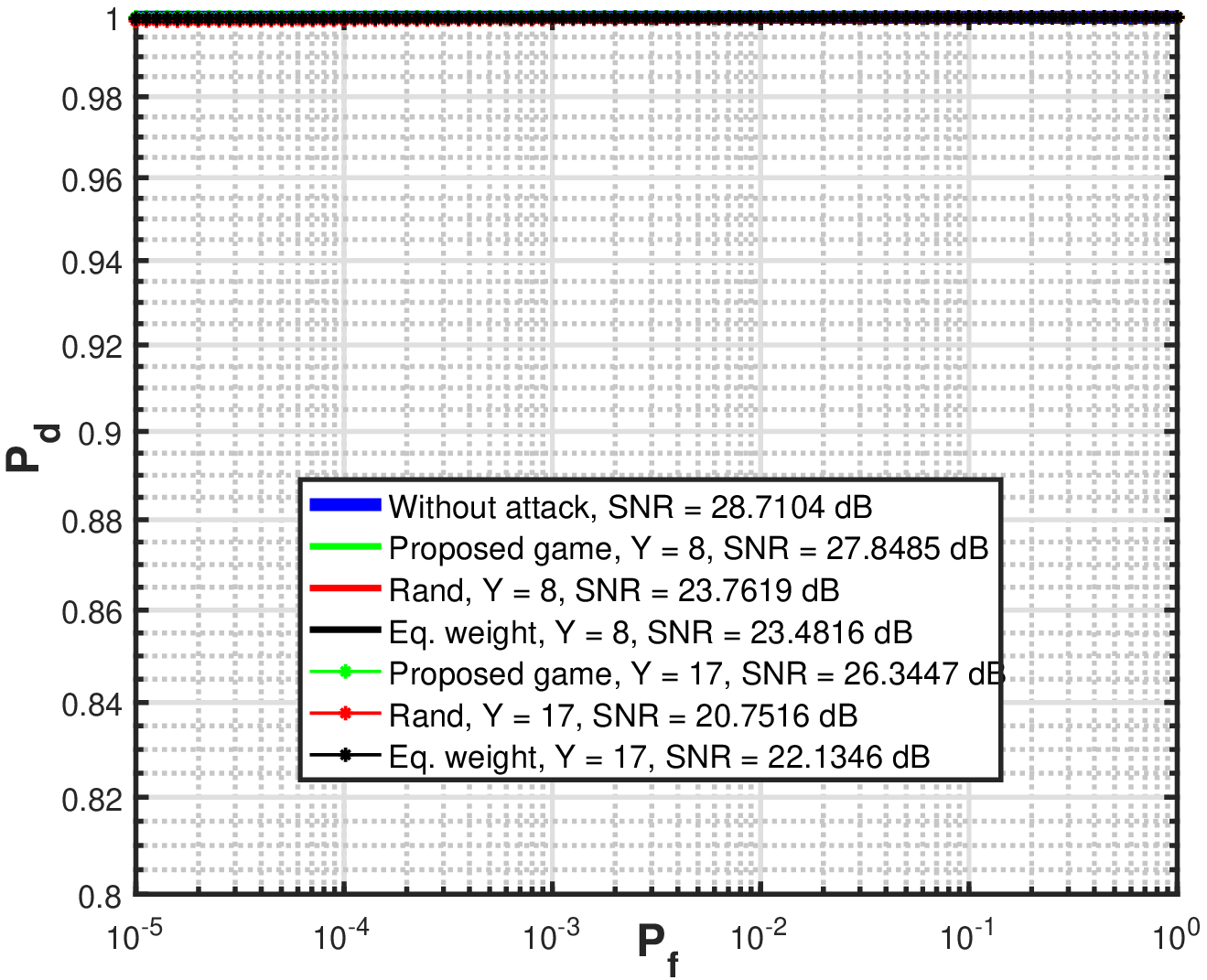} 
  \caption{IL}
  \label{fig:5}
\end{subfigure}\hfil 
\begin{subfigure}{0.3\textwidth}
  \includegraphics[width=\linewidth]{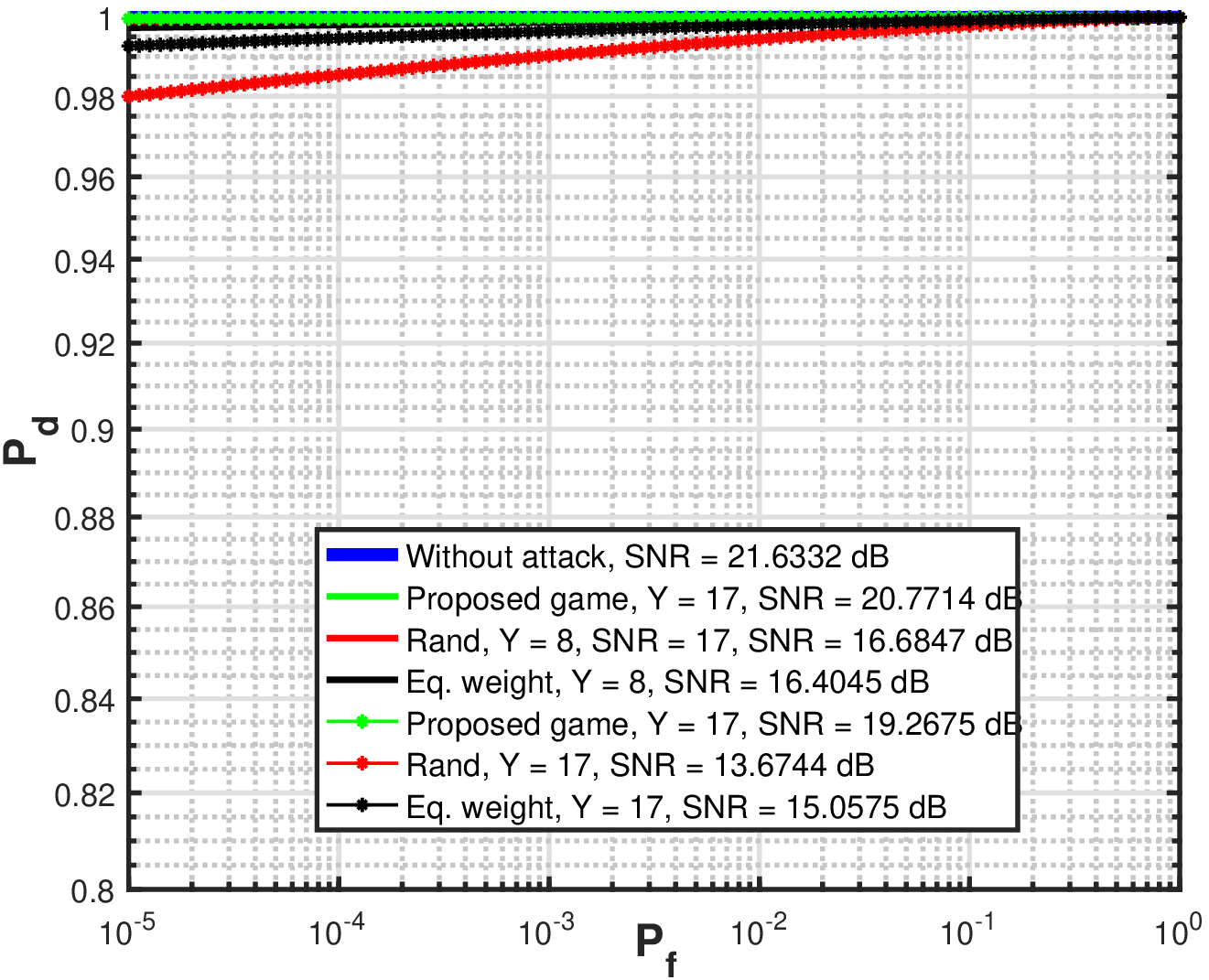} 
  \caption{IN}
  \label{fig:6}
\end{subfigure}
\caption{$P_d$ vs. $P_f$ of fluctuating PU with $Y = \{8,17\}$ over the six environments.}
\label{fig:images}
\end{figure*}

Figure \ref{fig:images} shows the presented study on the fluctuating PU performance ($P_d$ vs. $P_f$) over the six environments OL, ON, UL, UN, IL, and IN, respectively. Obviously, the results obtained are close to each other. It is clearly shown that, both OL, UL, and IL environments represents very close perfect performance. The ON environment presents the worst performance as compared to the other environments results because of being a harsh environment. This is a result of being the ON possesses the highest path loss exponent, which owns much more impact on the path loss, among the other environments. However, it is still a feasible performance.

\begin{figure}[!h]
\begin{center}
\includegraphics[width =\columnwidth, height = 10cm]{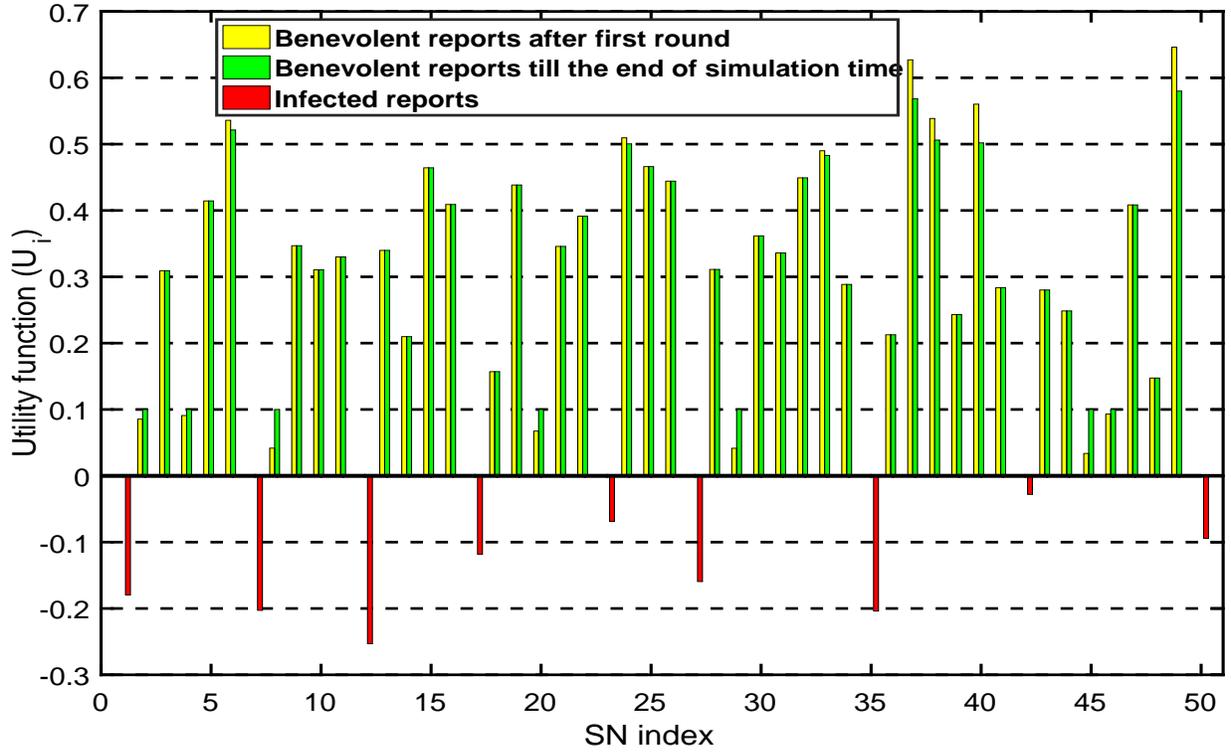}
\caption{Utility function values based on the proposed Stackelberg game.}
\label{utility}
\end{center}
\end{figure}

\section{Conclusion}\label{conclusion}
In this paper, we have presented an effective attack-defense strategy based on the Stackelberg game along with MF to alleviate the problem of corrupted SNs reports delivered at FC in WSNs-based CR over different six communication environments in which the data transmission is organized by TDMA technique. Moreover, TDMA is adopted to handle the complexity of MF along with the proposed model and to avoid the delivered reports collision. We also considered a realistic scenario of the SNs HW failure. However, the proposed model can detect this malfunction. Based on the obtained simulation results, the proposed model achieves effective protection against the SSDF malicious attack effect. In addition, the proposed approach protects about 83\% of the total number of SNs reports at the presence of the intelligent attack  manipulations.  Moreover, equilibrium can be achieved with minimal number of rounds. Therefore, the detection performance has been improved for our proposed approach, leading to an almost error-free spectrum access against the SSDF attack impact under the simulated scenarios. Furthermore, the number of packets that are correctly received by FC from the deployed SNs are promoted and the lost non-beneficial energy due to the attack impact has been managed. Consequently, the proposed Stackelberg game model, among the intelligent and adaptive game-theoretic approaches, along with MF and TDMA strictly prove beneficial for improving spectrum sensing, mitigating security threats, enhancing data privacy, and managing the power consumption in WSNs-based CR.

\begin{table*}[b]
\caption{Action matrix of the proposed Stackelberg game.}
\label{action_matrix}
\centering
\small
\begin{tabular}{cl|l|l|l|l|l|l|l|l|}
\cline{3-10}
\multicolumn{1}{l}{}                              &                  & \multicolumn{8}{c|}{Attack actions}                                                                                                                                                                                                    \\ \cline{2-10} 
\multicolumn{1}{l|}{}                             & Actions sequence & 1                 & 2                        & $\cdots$ &                                            &                                 & $\cdots$                                   & $\cdots$ & $\mathcal{|N|}$        \\ \hline
\multicolumn{1}{|c|}{\multirow{6}{*}{\begin{tabular}[c]{@{}l@{}}\\ \\ \\  FC\\ actions\end{tabular}}} & 1                & $x_1,y_1$         & $x_2,y_2$                & $\cdots$ & $x_{\gamma-1},y_{\gamma-1}$                & $x_{\gamma+1},y_{\gamma+1}$     & $\cdots$                                   & $\cdots$ & $x_{\mathcal{|N|}},y_{\mathcal{|N|}}$ \\ \cline{2-10} 
\multicolumn{1}{|c|}{}                            & 2                & $x^{'}_1,y^{'}_1$ & $x_2,y_2$                & $\cdots$ & $x_{\gamma-1},y_{\gamma-1}$                & $x_{\gamma+1},y_{\gamma+1}$     & $\cdots$                                   & $\cdots$ & $x_{\mathcal{|N|}},y_{\mathcal{|N|}}$ \\ \cline{2-10} 
\multicolumn{1}{|c|}{}                            & $\vdots$         & $\vdots$          & $\vdots$                 & $\vdots$ & $\vdots$                                   & $\vdots$                        & $\vdots$                                   &          &                                       \\ \cline{2-10} 
\multicolumn{1}{|c|}{}                            & $\vdots$         & $x^{'}_1,y^{'}_1$ & $x^{'}_2,y^{'}_2+\alpha$ & $\cdots$ & $x_{\gamma-1},y_{\gamma-1}$                & $x_{\gamma+1},y_{\gamma+1}$     & $\cdots$                                   & $\cdots$ & $x_{\mathcal{|N|}},y_{\mathcal{|N|}}$ \\ \cline{2-10} 
\multicolumn{1}{|c|}{}                            & $\vdots$         & $x^{'}_1,y^{'}_1$ & $x^{'}_2,y^{'}_2$        & $\cdots$ & $x^{'}_{\gamma-1},y^{'}_{\gamma-1}+\alpha$ & $x_{\gamma+1},y_{\gamma+1}$     & $\cdots$                                   & $\cdots$ & $x_{\mathcal{|N|}},y_{\mathcal{|N|}}$ \\ \cline{2-10} 
\multicolumn{1}{|c|}{}                            & $\mathcal{|N|}$  & $x^{'}_1,y^{'}_1$ & $x_2,y_2$                & $\cdots$ & $x_{\gamma-1},y_{\gamma-1}$                & $x^{'}_{\gamma},y^{'}_{\gamma}$ & $x^{'}_{\gamma+1},y^{'}_{\gamma+1}+\alpha$ & $\cdots$ & $x_{\mathcal{|N|}},y_{\mathcal{|N|}}$ \\ \hline
\end{tabular}
\end{table*}

%
%

\textbf{APPENDIX A}

$\because$ Attack strategy $AS = \\{y_1, y_2, \cdots, y_{\mathcal{|N|}} = \alpha x_1, \alpha x_2, \cdots, \alpha x_{\mathcal{|N|}}}$  

$\because$ the attacker benefit is maximized based on redistributing its budget from $y_r$ to $y_q$ to manipulate the defender, where ($r \neq q$).

\begin{equation*}
y_1, y_2, \cdots, y_r - \phi, y_q + \alpha, \cdots, y_{\mathcal{|N|}},
\end{equation*}
where the attacker deducts an attack budget $\alpha$ from the strong protected CM and add that to the weak protected CM. 

$\because$ The alternative attack by redistributing its budget and sorting those CMs in a descending order based on the previous status of the utility functions is denoted as

\begin{equation}
AS_A = {y_1^{'}, y_2^{'}, \cdots, y_{\mathcal{|N|}}^{'}}
\end{equation}

$\because$ The redistribution of the $AS$ is denoted as $D^\alpha_{r,q}$.\\

$\therefore$ any $AS_A$ can be represented by a redistribution sequence budget $\mathcal{|N|}-1$ from the optimal attack strategy ($AS^*$) as

\begin{equation*}
D_{1,2}^{\alpha_1}, D_{2,3}^{\alpha_2}, \cdots, D_{\mathcal{|N|}-1,\mathcal{|N|}}^{\alpha_{\mathcal{|N|}-1}}
\end{equation*}

Assume the budget redistribution from $AS_A \implies AS^*$ as

\begin{equation*}
D(1), D(2), \cdots, D(N-1) 
\end{equation*}

\noindent$\because$ The mathematical induction with respect to heuristic concept is utilized by $\alpha$-th redistribution for the $i$-th CM ($U_i$) in the ($WL$) depending on the first node (SCM) as 

\begin{equation}
U_1 = (y_1-L \cdot \alpha)- x_1,
\end{equation}
where $L_{>\mathbb{R}}$ is the subtracted number of ($\alpha$)'s from the strongest report budget given that the strongest report is the top of heap of the $(SL)$. More specifically, $U_1 > U_2$.


\begin{equation}
AS^* = max \; U_a \; \forall i = (1,2, \cdots, \mathcal{|N|})
\end{equation} 

\begin{equation}
\begin{aligned}
U_{i}\frac{\overset{SL}{>}}{\underset{WL}{
<}}\gamma, i \in ({1,2,...,\mathcal{|N|}}).
\end{aligned}
\end{equation}

To check the availability of finding $AS_A > AS^*$

\begin{equation}
U_{a_i} > U_{d_i}|i \notin WL \; \text{or} \; SL
\end{equation} 

\noindent$\because$ The attacker concentrates on $WL$.\\ \\
$\therefore$ There is no $AS_A$ satisfies $U_{a}> U_{d} \; \forall i$ as illustrated below using the chain budget redistribution.  

\begin{equation*}
\begin{split}
AS^* = & {y_1, y_2, \cdots, y_{\gamma-1}, y_{\gamma+1}, \cdots, y_{|\mathcal{N}|}}\\
& D_{1, 2} \Downarrow \\
& {y^{'}_1, y_2+{\alpha}, \cdots, y_{\gamma-1}, y_{\gamma+1}, \cdots,y_{|\mathcal{N}|}}\\
& D_{2, 3} \Downarrow \\
& {y^{'}_1, y_2^{'}, \cdots, y_{\gamma-1}, y_{\gamma+1}, \cdots, y_{|\mathcal{N}|}}\\
& \vdots \\
& D_{\gamma-1+\alpha, \gamma} \Downarrow \\
& {y^{'}_1, y_2^{'}, \cdots, y_{\gamma-1+\alpha}, y_{\gamma}, y_{\gamma+1}, \cdots, y_{|\mathcal{N}|}}\\
\end{split}
\end{equation*}

$\therefore$ the action matrix leading to the NE presented in Eq. \ref{NE} is given by Table \ref{action_matrix}.

\bibliographystyle{IEEEtran}
\bibliography{ref}

\end{document}